\documentclass[11pt, oneside]{article}   	% use "amsart" instead of "article" for AMSLaTeX format
\usepackage{geometry}                		% See geometry.pdf to learn the layout options. There are lots.
\usepackage{graphicx}				% Use pdf, png, jpg, or eps with pdflatex; use eps in DVI mode
\usepackage{amssymb}
\usepackage{amsmath}
\usepackage{subfig}
\usepackage{caption}
\usepackage{comment}

\usepackage[utf8]{inputenc}
\usepackage{amsmath}
\numberwithin{equation}{section}
\begin{document}
\title{Quantum Computing for \\
Inflationary, Dark Energy and Dark matter Cosmology }
\author{ Amy Joseph$^1$, Juan-Pablo Varela$^2$,  Molly P. Watts$^3$, Tristen White$^4$,\\ Yuan Feng$^5$, Mohammad Hassan$^6$, Michael McGuigan$^7$,\\   \footnote{ (1) Arizona State University, (2) University at Albany (SUNY) current address U. Mass. Boston, (3) Columbia University, (4) Juniata College, (5) Pasadena City College, (6) The City College of New York, (7) Brookhaven National Laboratory,   }
}
\date{}
\maketitle
\begin{abstract}
Cosmology is in an era of rapid discovery especially in areas related to dark energy, dark matter and inflation.  Quantum cosmology treats the cosmology quantum mechanically and is important when quantum effects need to be accounted for, especially in the very early Universe. Quantum computing is an emerging new method of computing which excels in simulating quantum systems. Quantum computing may have some  advantages when simulating quantum cosmology, especially because the Euclidean action of gravity is unbounded from below, making the implementation of Monte Carlo simulation problematic. In this paper we present several examples of the application of quantum computing to cosmology. These include a dark energy model that is related to Kaluza-Klein theory, dark matter models where the dark sector is described by a self interacting gauge field or a conformal scalar field and an inflationary model with a slow roll potential. We implement  quantum computations in the IBM QISKit software framework and show how to apply the Variational Quantum Eigensolver (VQE) and Evolution of Hamiltonian (EOH)  algorithms to solve the Wheeler-DeWitt equation that can be used to describe the  cosmology in the mini-superspace approximation. We find excellent agreement with classical computing results and describe the accuracy of the different quantum algorithms. Finally we  discuss how these methods can be scaled to larger problems going beyond the mini-superspace approximation where the quantum computer may exceed the performance of classical computation.

%In this paper we discuss the inflationary cosmology  on a quantum computer. %We discuss several variants of inflationary cosmology including dilaton %cosmology, axion cosmology and Kaluza-Klein Cosmology. We also discuss the %coupling to fermions, gauge fields and the Higgs field. First we discuss 5how to set up the various Hamiltonians on the quantum computer using %various Hamiltonian Mappings to qubits. Then we couple these theories to %gravity and study the Hamiltonian constraint or Wheeler-DeWitt equation on %the quantum computer treating the system using the Variationa Quantum %Eigensolver (VQE) quantum algorithm.This can be used to estimate the wave %function of the Universe that can serve as an initial condition for early %Universe evolution in inflationary cosmology determined on near term %quantum computers. We also discuss inhomogeneous  cosmologies and how these %may lead to a quantum advantage of classical computations for these type of %models.

\end{abstract}
\newpage

%1.Introduction

%2.Quantum computing 

%3.Inflationary Cosmology

%4.Dark Energy Cosmology

%5.Dark Matter Cosmology

%6.Quantum Cosmology

%7.Conclusion

%Appendix A. Other forms of potentials

\section{Introduction}

Three important areas of modern cosmology are inflationary cosmology, dark energy cosmology and dark matter cosmology. In inflationary cosmology the very early Universe is described by an inflationary phase where the Universe expanded exponentially. The type of inflationary model which fits best with experiment for example from the cosmic microwave background measurements is slow roll inflation where the potential of the inflaton scalar field has a long flat region. The particular inflationary potential we will consider in this paper is the Starobinsky potential which fits well with existing data.

Dark energy is the energy density of space itself and  describes the driving mechanism for the accelerating expansion of the universe. It's a repulsive force, acting against the attractiveness of gravity, which contrasts our initial assumptions of a static universe. It's considered to be a constant value (contrary to quintessence). One result in the exploration of dark energy, is the discrepancy between the theoretical value and the observed value; being around 120 orders of magnitude off, this inconsistency is considered one of the greatest puzzles in physics. It's been measured to have an astonishingly small value (on the order of $10^{-120}$ in Planck units). One model of dark energy which we consider in this paper involves Kaluza-Klein theory which  incorporates  higher dimensions with a much larger value of the cosmological constant or $\Lambda_{D+4}$ where  $D$ is the number of extra dimensions. This method involves a big vacuum energy that is “hidden” in  the extra dimensional lambda. The bulk of the vacuum energy is used  to hold the extra dimensional space in a static state  through a delicate balance with its internal curvature.  The extra vacuum  energy which is left over can be arbitrarily small and is what is perceived in this model as a small four dimensional cosmological constant. 

Dark matter makes up the vast majority of matter in the Universe and plays an important role in galaxy formation. Dark matter interacts with gravity so one can consider dark matter models of cosmology. These models depend on what makes up dark matter of which there a number of candidates. In this paper we discuss two such candidate models where the quantum field associated with dark matter consists of a conformally coupled scalar field and another model where the dark matter field consists of a self interacting gauge field where the gauge group lives outside the gauge group of the standard model.

Finally one can consider quantum cosmology where one treats gravity quantum mechanically and cosmology treats the evolution of the Universe through the Feynman path integral or Hamiltonain formulations. The particular approach we will consider in this papers is the Wheeler-DeWitt equation which is related to the Hamiltonian constraint of the cosmology. 

All of these cosmologies can be considered using classical computing but there may be some advantages through considering them in the context of quantum computing \cite{Kocher:2018ilr}
\cite{Ganguly:2019kkm}
\cite{Li:2017gvt}
\cite{Mielczarek:2018nnd}
\cite{Mielczarek:2018jsh}
\cite{Li:2020kbv}
\cite{Antonini:2019qkt}
\cite{Brown:2019hmk}
\cite{Nezami:2021yaq}\cite{Mocz:2021ehj}. This is  especially the case  if the number of degrees of freedom and hence the dimension of Hilbert space becomes large or if a purely Lorentzian approach is favored, which may preclude the use of the Monte Carlo approach of computing path integrals \cite{Berger:1993fm}.

This paper is organized as follows. In section 2 we give a brief introduction to some of the quantum computing techniques we will use in the paper. In section 3 we will discuss that application of quantum computing to the Hamiltonian describing the Starobinsky potential of inflationary cosmology. In section 4 we discuss the Hamiltonian of a theory of dark energy cosmology involving Kaluza-Klein theory with extra dimensions and a large value of the higher dimensional cosmological constant. We discuss two models, one where the internal radius is small and is described by a single radius and another model where there are two internal radii. In section 5 we shall follow a similar approach to the study of the Hamiltonian formulation of two models of dark matter cosmology involving either conformally coupled scalar fields or self interacting gauge fields.  Finally in section 6 we discuss the application of quantum computing to quantum cosmology using the minisuperspace approximation and in section 7 we discuss the main conclusions of the paper.

\section{Quantum Computing}

Quantum computing is a developing computing platform based on the principles of quantum mechanics. Based on the difficulty of simulating quantum systems on classical computers Feynman suspected that if quantum computers could be built they would be more efficient at simulating quantum systems than classical computers because of the large dimension of the Hilbert spaces which were involved. There are several quantum algorithms that can be applied to mathematical physics problems each of which offers a potential speedup. In particular the Variational Quantum Eigensolver (VQE) \cite{Kandala} and Evolution of Hamiltonian (EOH) algorithms will be  applied  to cosmological problems in this paper. 

The VQE is a quantum variational method where expectation values  of a Hamiltonian are measured on a quantum computer in a variational wave form that depends on parameters such as $ \frac{{\left\langle {\psi ({\theta _i})} \right|H\left| {\psi ({\theta _i})} \right\rangle }}{{\left\langle {\psi ({\theta _i})} \right|\left. {\psi ({\theta _i})} \right\rangle }}$ that are optimized on an iterative basis on a classical computer. Thus the VQE is a hybrid algorithm as parts of the algorithm are run on a quantum computer in conjunction with other parts, in particular the optimization of parameters, that run on a classical computer. 

The EOH represents the unitary evolution operator $ U(t) = {e^{ - iHt}} $ in terms of quantum gates on a quantum computer. It take an initial state represented in terms of qubits and evolved it to a final state by computing the quantity $K(i,f;t) = \left\langle {{\psi _i}} \right|{e^{ - iHt}}\left| {{\psi _f}} \right\rangle $. As this quantity can be represented as a Lorentzian Path Integral this algorithm illustrates the potential advantage of quantum computing as only Euclidean path integrals can be efficiently done on a classical computer using the Monte-Carlo method. The quantum computer accomplishes this by mapping the path integral to be computed to another quantum system which does its own version of the path integral native to the physical realization of the quantum computer such as a superconducting circuit, an ion trap or nuclear spin. The answer is then translated into the time evolution of the state for the path integral that we want to compute. Thus one never needs to know how to compute the path integral directly, but let's nature do the evolution indirectly on the quantum computer.

Whichever method one uses the first step is to make a Hamiltonian mapping onto qubits which usually means choosing a basis to represent the operators in the system. 
 The first step is to represent the Hamiltonian as an $N \times N $ matrix using a discrete quantum mechanics approximation to the quantum mechanical operators which would be infinite dimensional for bosonic observables \cite{Miceli}\cite{Okock}\cite{Korsch}\cite{Motycka}. In this paper we will use three different types of discrete Hamiltonians bases described below.

\subsection*{Gaussian or Simple Harmonic Oscillator basis}

This is a very useful basis based on the matrix treatment of the simple harmonic oscillator which is sparse in representing the position and momentum operator. For the position operator we have:
\begin{equation} 
 X_{osc} = \frac{1}{\sqrt{2}}\begin{bmatrix}
 
   0 & {\sqrt 1 } & 0 &  \cdots  & 0  \\ 
   {\sqrt 1 } & 0 & {\sqrt 2 } &  \cdots  & 0  \\ 
   0 & {\sqrt 2 } &  \ddots  &  \ddots  & 0  \\ 
   0 & 0 &  \ddots  & 0 & {\sqrt {N-1} }  \\ 
   0 & 0 &  \cdots  & {\sqrt {N-1} } & 0  \\ 
\end{bmatrix}
  \end{equation}
while for the momentum operator we have:
\begin{equation}
 P_{osc} = \frac{i}{\sqrt{2}}\begin{bmatrix}
 
   0 & -{\sqrt 1 } & 0 &  \cdots  & 0  \\ 
   {\sqrt 1 } & 0 & -{\sqrt 2 } &  \cdots  & 0  \\ 
   0 & {\sqrt 2 } &  \ddots  &  \ddots  & 0  \\ 
   0 & 0 &  \ddots  & 0 & -{\sqrt {N-1} }  \\ 
   0 & 0 &  \cdots  & {\sqrt {N-1} } & 0  \\ 
\end{bmatrix}
  \end{equation}
%The  Morse  Hamiltonian $H_{-}$ is then 
%\begin{equation}
% H_{-}=P_{osc}^2 + Exp(-2X_{osc}) - (2A+1) Exp(-X_{osc}) + A^2 I   
%\end{equation}
%where $Exp$ refers to the Matrix exponential and $I$ is the $N \times N$ identity matrix.

\subsection*{Position basis}

In the position basis the position matrix is diagonal but the momentum matrix is dense and constructed from the position operator using a Sylvester matrix $F$. In the position basis the position matrix is:
\begin{equation}
{\left( {{X_{pos}}} \right)_{j,k}} = \sqrt {\frac{{2\pi }}{{4N}}} (2j - (N + 1)){\delta _{j,k}}
\end{equation}
and the momentum matrix is:
\begin{equation}{P_{pos}} = {F^\dag }{X_{pos}}F\end{equation}
where 
\begin{equation}{F_{j,k}} = \frac{1}{{\sqrt N }}{e^{\frac{{2\pi i}}{{4N}}(2j - (N + 1))(2k - (N + 1))}}\end{equation}
%The Morse Hamiltonian is  formed from
%\begin{equation}
 %H_{-}=P_{pos}^2 + Exp(-2X_{pos}) - (2A+1) Exp(-X_{pos}) + A^2 I   
%\end{equation}
%but in this case the matrix exponential is very simple as it is the exponential of a diagonal matrix.

\subsection*{Finite difference basis}

This is the type of basis that comes up when realizing differential equations in terms of finite difference equations. In this case the position operator is again diagonal but the momentum operator although not diagonal is still sparse. In the finite difference basis the position matrix is:
$${\left( {{X_{fd}}} \right)_{j,k}} = \sqrt {\frac{1}{{2N}}} (2j - (N + 1)){\delta _{j,k}}$$
and the momentum-squared matrix is:
\begin{equation} 
 P_{fd}^2 = \frac{N}{2}\begin{bmatrix}
 
   2 & - 1  & 0 &  \cdots  & 0  \\ 
   -1 & 2 & -1 &  \cdots  & 0  \\ 
   0 & -1 &  \ddots  &  \ddots  & 0  \\ 
   0 & 0 &  \ddots  & 2 & -1  \\ 
   0 & 0 &  \cdots  & -1 & 2  \\ 
\end{bmatrix}
  \end{equation}
%The Morse Hamiltonian is then:
%\begin{equation}
% H_{-}=P_{fd}^2 + Exp(-2X_{fd}) - (2A+1) Exp(-X_{fd}) + A^2 I   
%\end{equation}
Whatever basis one uses, one needs to map the Hamiltonian to a an expression in terms of a sum of tensor products of Pauli spin matrices plus the identity matrix which are called Pauli terms. As there are four such matrices, the maximum number of terms in this expansion is $4^n$  where $n$ is the number of qubits. In most of our simulations the number of qubits was fixed at 4 so that the maximum number of Pauli terms was 256.

Finally if one has more than one bosonic field in the theory on can form the Hamiltonian for the multi-boson theory by using tensor products. For example for a two boson system we have:
\[\begin{array}{l}
{X_1} = X \otimes I\\
{X_2} = I \otimes X\\
{P_1} = P \otimes I\\
{P_2} = I \otimes P
\end{array}\]
Then one computes the Hamiltonian of the multi-boson system by taking functions of these operators. Because of the tensor product, the size of the multi-boson Hamiltonian grows expontentially  with the number of bosons. So if each boson was represented as a $16 \times 16 $ matrix, a two boson system would prepresnt the Hamiltonian as  a $256 \times 256 $ matrix and a 15 boson Hamiltonian would require representing the Hamiltonian as a $16^{15} \times 16^{15} $ matrix which would require a quantum computer with 60 qubits.

\section{Inflationary Cosmology}

We began by reviewing the Friedmann equation and the Starobinsky potential in order to create some groundwork for the description of quantum cosmology in the Hamiltonian formulation.

\subsection*{The Friedmann Equations}

Considered to be one of the most important equations, the Friedmann equation describes the expansion of the Universe. It chronicles the life of the Universe, predicting where it began and where it will end up, simply by inputting a few crucial values. Here, the Friedmann equation is written as,
\begin{equation}\frac{\dot{a}^2 + kc^2}{a^2} = \frac{8 \pi G \rho + \Lambda c^2}{3}\end{equation}
where $a(t)$ is the scale factor (radius) of the Universe at time $t$, $k$ is the curvature of the Universe (defined as $-1$, $0$, or $1$ for closed, flat, and open Universe, respectively), $G$ is the gravitational constant, $\rho$ is the energy/matter density ($\frac{C}{a^4}$ for radiation, $\frac{m}{a^3}$ for matter, respectively), and $\Lambda$ is the cosmological constant. With this equation one can examine a variety  of cosmological models for a number of different scenarios. The Friedmann equations can be derived from the simple Lagrangian.
\begin{equation}L = c\left[ { - 3\frac{{a{{\dot a}^2}}}{N} + 3Nka - \Lambda {a^3}N + {a^3}\frac{1}{2}\frac{{{{\dot \phi }^2}}}{N} - N{a^3}V(\phi )} \right]\end{equation}
Varying with respect to $N$, $a$ and $\phi$ yeild the Freidmann equations coupled to a single scalar field.
\begin{align}
 - 3\frac{{{{\dot a}^2}}}{{{a^2}}} &- 3\frac{k}{{{a^2}}} + \Lambda  + \frac{1}{2}{{\dot \phi }^2} + V(\phi ) = 0\nonumber\\
 - 2\frac{{\ddot a}}{a} &- 2\frac{{{{\dot a}^2}}}{{{a^2}}} + \frac{k}{{{a^2}}} - \Lambda  + \frac{1}{2}{{\dot \phi }^2} - V(\phi ) = 0r\\
 \ddot \phi  &+ 3\frac{{\dot a}}{a}\dot \phi  + \frac{{dV}}{{d\phi }} = 0\nonumber
\end{align}
%\begin{figure}[htp]
 %   \centering
 %   \includegraphics[width=6cm]{cdm.jpg}
 %   \caption{Plot of the Benchmark: $\Lambda$CDM model, a fitting theory %for the Universe we observe. This particular model has parameters: %$k=C=0$, $\Lambda=0.73$, $m=0.27$, baryonic matter $= 0.04$, and dark %matter $= 0.23$.}
  %  \label{fig:cdm}
%\end{figure}
These equation represent the generalization of the Friedmann equations to case of a single scalar field \cite{NASA/WMAP}\cite{Wiltshire:2003} \cite{Valentino} 
\cite{Brizuela:2016gnz}
\cite{Garay:1990re}
\cite{Halliwell:1986ja}
\cite{Ha}
\cite{Atkatz}
\cite{deAlwis:2018sec}
\cite{Domingos}and we will apply them to the case of a particular slow roll inflation model below.

\subsection*{The Starobinsky Potential}

The Starobinsky potential is a model for cosmological inflation. It is a type of scalar potential which closely resembles the Morse potential from chemistry - the noticeable difference being that the Starobinsky is reflected over the vertical axis. The Starobinsky potential is given by
$$V(\phi)=M^4_1 (1-e^{\frac{\phi}{M_{2}}})^2$$
where $\phi$ is the scalar field for the potential and $M_1, M_2$ are parameters.  A  plot of the Starobinsky potential is found in figure 1 for $M^4_1 = 29.167$ and $M_2 = 7.638$ in Planck units. A plot of the solution for the inflaton field is plotted in figure 2. The inflaton field slowly rolls from large negative values a then oscillates near the bottom of the potential eventually approaching zero.

\begin{figure}
    \centering
    \begin{minipage}[b]{0.7\linewidth}
      \centering
      \includegraphics[width=\linewidth]{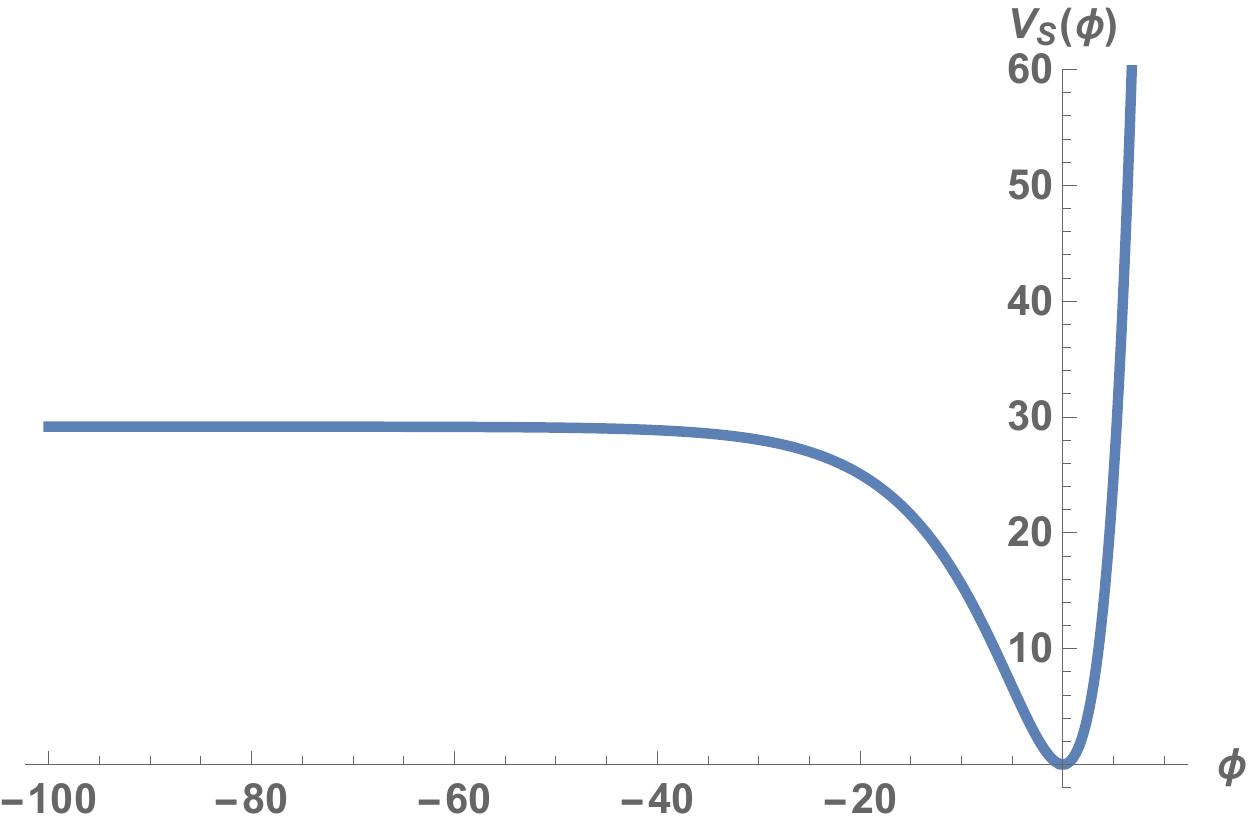}
    \end{minipage}
    \caption{Plot Starobinsky potential used in inflationary cosmology. }
    \label{3-qubit_unit}
\end{figure}
%\begin{figure}[htp]
 %   \centering
 %   \includegraphics[width=7cm]{star.jpg}
 %   \caption{Plot of the Starobinsky potential as a function of the scalar %field for the potential. Note the resemblance to the Morse potential %simply reflected across the vertical axis.}
 %   \label{fig:star}
%\end{figure}

%This potential strongly mimics the shape of the dark energy potential, making it an incredibly useful building block. From here, we were able to explore the dark energy potential using classical and quantum computing.
\begin{figure}
    \centering
    \begin{minipage}[b]{0.5\linewidth}
      \centering
      \includegraphics[width=\linewidth]{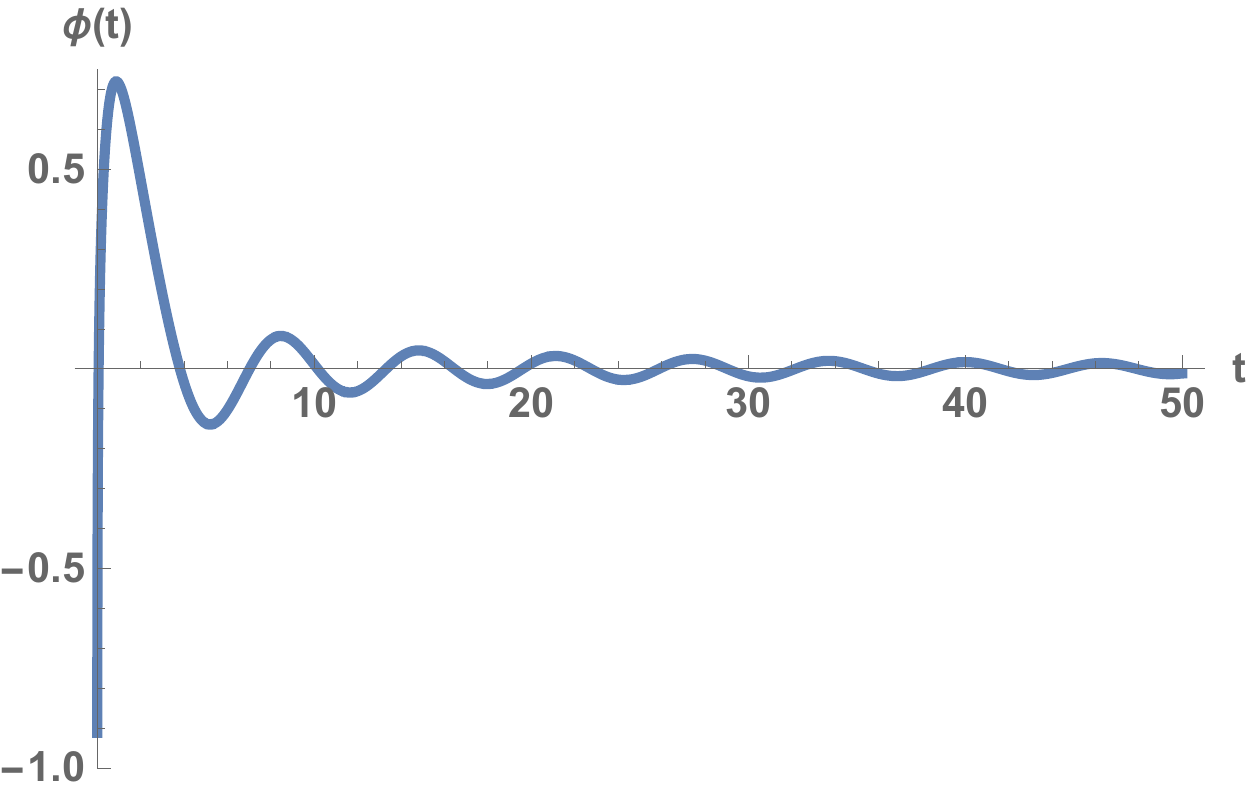}
    \end{minipage}
    \caption{Time evolution of the Inflaton field $\phi$ in the Starobinsky potential. }
    \label{3-qubit_unit}
\end{figure}

\newpage

%As the Starobinsky potential is equivalent to the Morse potential (except for a reflection about the y axis) whose energy eigenstates and eigenfunctions are exactly known we can write down the exact energies for the Starobinsky potential as a function of $\mu_1$ and $\mu_2$. Also like in the recent  the treatment of the Morse potential these ground state energies and wave functions can be computed using the VQE quantum algoritm.  algorithm.

\subsection*{Quantum computing for the Starobinsky potential}

Because the bottom of the Starobinsky potential resembles a simple Hamonic oscillator potential we expect a shift in the ground state energy of the inflaton field due to quantum fluctuations of its zero mode. Also particle excitations in the inflaton field can cause a shift in the ground state energy. Thus we parametrize $E_0 = V_{min} + \Lambda_{quant}$ where $\Lambda_{quant}$ represents the shift in the ground state energy to quantum effects.

%To calculate the ground state energy, we construct Hamiltonian as 16 by 16 matrix, which means it requires 4 qubits to do the quantum computation. We used 135 pauli terms. Using the  VQE we obtained the upper bound on the ground state energy to 0.49789949, with program running time as 1.5408587455749512s. This compared well to the  to exact result of 0.49785652.
%For $\mu_1^4 = 29.1667$ and $\nu_2 = 7.6376 $ we find the exact ground state energy $E_0 = .47857$. For the VQE calculation with 4 qubits and the harmonic oscillator basis we find

To calculate the ground state energy, we construct the Hamiltonian as 16 by 16 matrix using the oscillator basis described above, which means it requires 4 qubits to do the quantum computation. We used 135 Pauli terms, minimizing the Hamilitonian expectation value using one uses an optimization program running on a classical computer. Four of the optimizers available in QISKit are  Sequential Least SQuares Programming optimizer (SLSQP), Constrained Optimization By Linear Approximation optimizer (COBYLA), Limited-memory BFGS Bound optimizer (L-BFGS-B), and the Nelder-Mead optimizer. Using the  VQE we obtained the upper bound on the ground state energy as 0.4978995 in Planck units with the SLSQP optimizer with program running time as 1.54s. Compared to exact result as 0.49785652, there is 0.00863\% percentage error.
\begin{figure}[htp]
    \centering
    \includegraphics[scale=0.3]{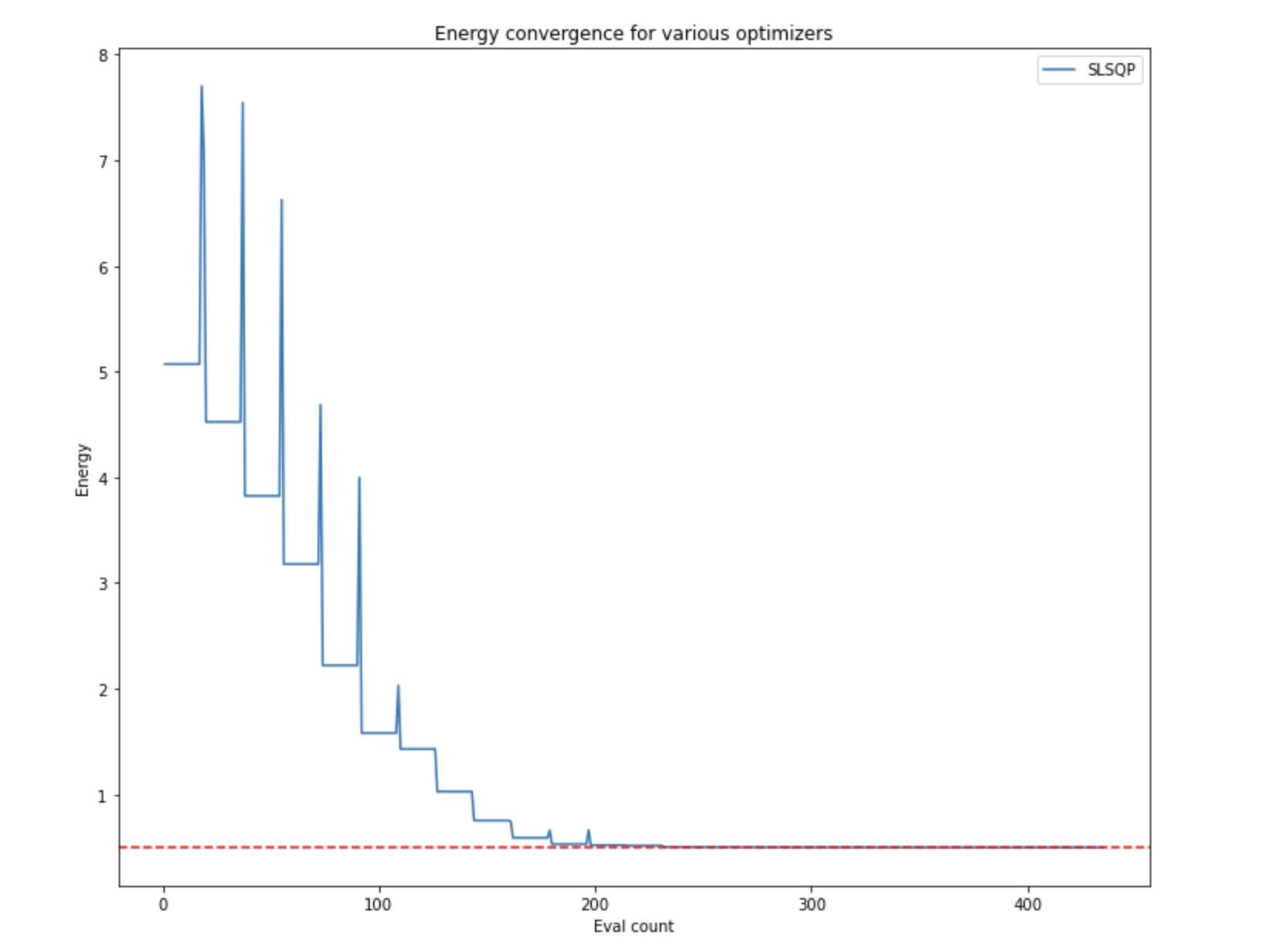}
    \caption{VQE energy convergence graph of Starobinsky potential using the SLSQP optimizer.}
    \label{fig:mesh1}
\end{figure}
\begin{table}
\begin{center}
\begin{tabular}{ |c|c|c|c|c|c| } 
 \hline
Model & Qubits & Pauli Terms & Exact  & VQE   \\
\hline
 
 Starobinsky Inflationary Potential & 4 & 135 & 0.49785652 & 0.4978995\\ 
 \hline
\end{tabular}
\end{center}
\caption{Results from the VQE computation for the Starobinsky inflationary potential using the SLSQP optimizer and 4  qubits.}
\end{table}
This is in excellent agreement with the exact value as shown in figure 3 and table 1. As $V_{min} = 0$ for the Starobinsky potential the nonzero value of $E_0$ would indicate that the Universe would undergo an addition exponential inflation after the initial inflation at a rate larger than what we expect from Dark Energy. One way to correct for this is is to subtract $\Lambda_{quant}$ from the Starobinsky potential. When this is done the Universe will still enter a initial inflationary period although with slightly lower effective cosmological constant $V_{subtracted}(\phi) = V_{original}(\phi) - \Lambda_{quant}$. In the next section we will discuss Dark Energy potentials which can more adequately account for the residual dark energy which becomes important at later times.

\section{Dark Energy Cosmology}

The size of the cosmological constant $\Lambda_4$ in present day is bounded by the dark energy maeasurements and is of the order $10^{-120}$ in Planck units \cite{Peebles}. Although it is puzzling and quite difficult to create a model universe that has such a small value for $\Lambda_4$, there are some plausible theories. One of these theories, which we will be using here, involves  a higher dimensional cosmological constant lambda, $\Lambda_{D+4}$ where $D$ is the number of extra dimensions. The value of $\Lambda_{D+4}$  is significantly larger than the four dimensional lambda, $\Lambda_4$, we perceive  \cite{Maloney}
\cite{Font:2002pq}
\cite{Brown:2014sba}
\cite{Brown:2013mwa}
\cite{Brown:2013fba}.

\subsection*{Single radius dark energy potential}
In order to model the dark energy potential, and incorporate higher dimensions and we use a dark energy potential of the form:
%$$V(\phi)=e^{\frac{4}{d-2}\phi}(a-be^{\phi}+ce^{2\phi})$$
\begin{equation}V(b) = {b^{ - D}}\left( {\frac{{Q_D^2}}{{{b^{2D}}}} - \frac{k}{{{b^2}}} + {\Lambda _{D + 4}}} \right)\end{equation}
Here $b$ is the radius of the extra $D$ dimensions. The different terms in the potential represent energy from $D$ form flux, curvature of the extra dimension and a higher dimensional cosmological constant from left to right. In this paper we choose $D=4$ so we have four extra dimensions supported by a four form  flux $Q_4$. Defining $\phi = \log b$ the dark energy Hamiltonian becomes:
\begin{equation}H = \frac{{p_\phi ^2}}{2} + {e^{ - 4\phi /c}}\left( {Q_4^2{e^{ - 8\phi /c}} - k{e^{ - 2\phi /c}} + {\Lambda _8}} \right)\end{equation}
We plot the dark energy potential for $Q_4^2 = 29.166$, $k = 58.33$, $c=45.84 $ and $\Lambda_8 = 34.55$ in Planck units in figure 4
%where $a$, $b$, $c$, and $d$ are all constant parameters consisting of defining features of the Universe. A series expansion provides the values for these terms %and creates the following dark energy potential: $a=1=$ curvature term, $b=2$, and $c=1.1845575279846237=\Lambda_8$,
%$$V(\phi)=\frac{7}{48(0.005)} \left( {e^\frac{-2\phi}{7.637626158259733}} -2{e^\frac{-\phi}{7.637626158259733}} + %{1.1845575279846237e^\frac{-2\phi}{7.637626158259733}} \right)$$
including its determined ground state energy value $\Lambda_4$. Because of the uncertainty principle the canonical momentum $p_{\phi}$ and $\phi$ cannot both be zero so there is a small positive shift from the minimum the potential due to the kinetic energy. For the parameters above we find the ground state energy  $\Lambda_4= 1.11637 \times {10^{ - 6}} M_{Planck}^{-4}$. This is still much larger than the observed value of $\Lambda_4 \approx {10^{-120}} M_{Planck}^{4}$ but much reduced from the higher dimensional cosmological constant $\Lambda_8 = 34.5 5M_{Planck}^{4}$. In the next subsection we will describe how one can use the quantum computer and variational methods to set an upper bound on $\Lambda_4$ and describe the method that can be used for other dark energy potentials.

\begin{figure}[!htb]
    \centering
    \includegraphics[width=8cm]{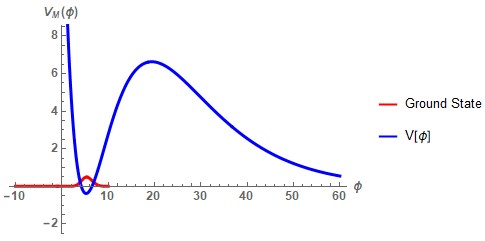}
    \caption{Plot of V[$\phi$] as a function of $\phi$, depicting the dark energy potential (blue) and the ground state energy for the potential (red). The red line is the value of the four dimensional cosmological constant, $\Lambda_4$. The bottom of the potential well is the classical value which gets raised by quantum effects. The value for the four dimensional cosmological constant was found to be $\Lambda_4 = 1.11637 \times {10^{ - 6}} M_{Planck}^{4}$ compared to  the actual value that is approximately $\Lambda_4 \approx {10^{-120}} M_{Planck}^{4}$. Although still far too large it is heading in the right direction.}
    \label{fig:DEP}
\end{figure}

\begin{figure}[htp]
    \centering
    \includegraphics[scale=0.3]{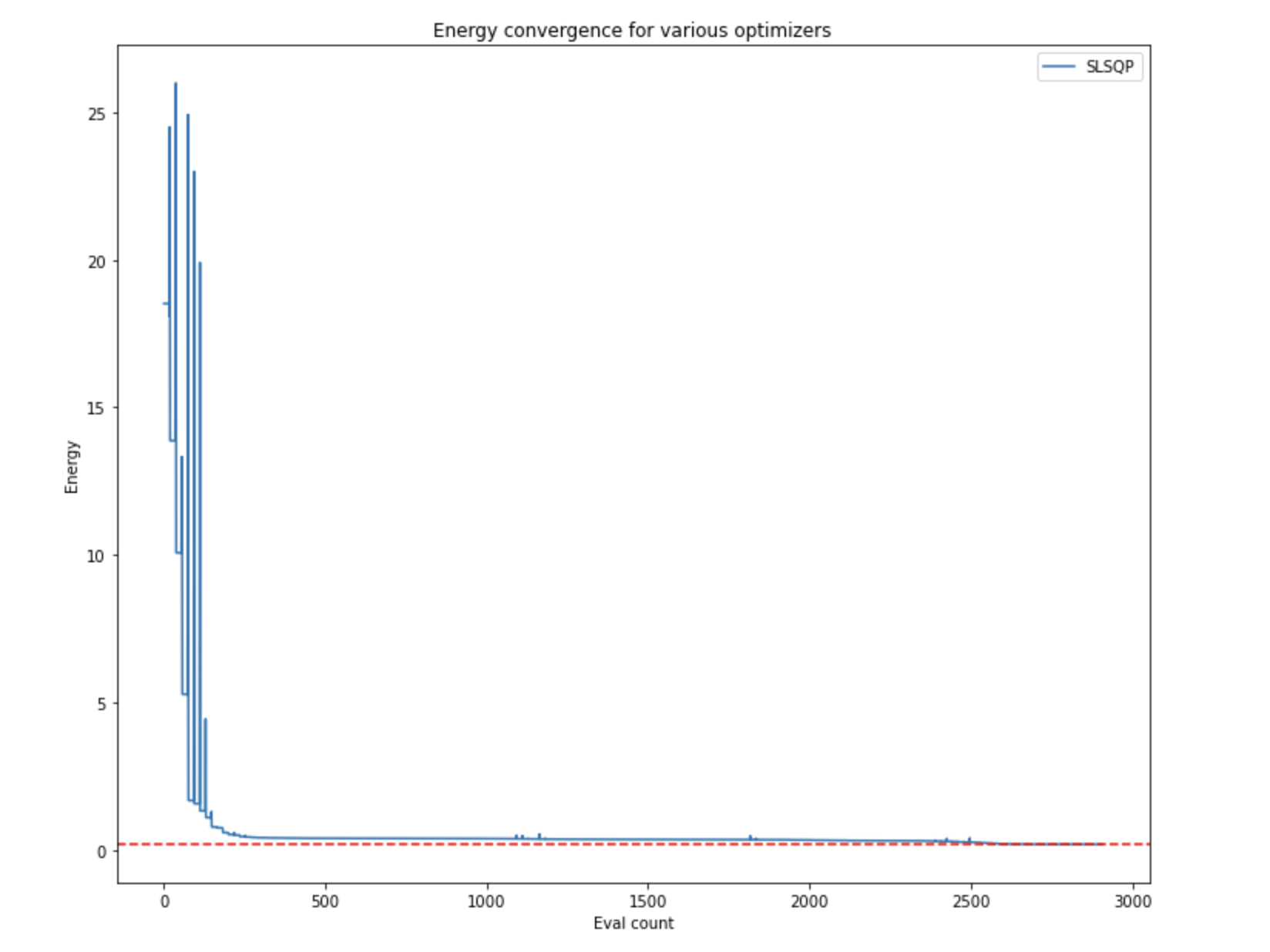}
    \caption{VQE energy convergence graph of single radius dark energy potential}
    \label{fig:mesh1}
\end{figure}

\subsection*{Metastable Vacua}

Recall the dark energy potential from Figure 3 above, and how it decreases as $\phi$ increases (dissimilar to the Starobinsky potential which plateaued). This is a key feature that leads to the conclusion that the dark energy potential is metastable. In essence, this means that the potential is stable for a certain length of time.  Therefore, there is a probability that a Universe sitting at the bottom of the potential well can tunnel out through the barrier. When this happens the Universe will decompactify and become higher dimensional. Thus, one must check that the time required for this to happen is longer than the age of the Universe for the dark energy model to be compatible to what we see. We can use the methods of \cite{Linde:1981zj}
\cite{Frieman:1985xs}
\cite{Kolb:1986nj}
\cite{Kolb:1987dd}
to do the necessary calculation. 
First, starting with a series expansion of the dark energy potential about it's minimum we find it is of the form,
\begin{equation}V(\phi ) = {V_{\min }} + \frac{{{M^2}}}{2}{\phi ^2} - \frac{\delta }{3}{\phi ^3} +  \ldots \end{equation}
where $V_{min} = -0.378498 $, $M^2 = 0.584376$ and $\delta = 0.139707$, and these are defining qualities of the expansion ($M$ is the frequency at the bottom of the well, whereas $\delta$ is taking control at the top of the well). The analysis of \cite{Linde:1981zj}
\cite{Frieman:1985xs}
\cite{Kolb:1986nj}
\cite{Kolb:1987dd} shows that for potentials of this form the tunneling action is given by,
\begin{equation}S_E = 205\frac{{{M^2}}}{{{\delta ^2}}}\end{equation}
And the tunneling probability for the potential is then
\begin{equation}\frac{S_E}{4}=51.25\frac{M^2}{\delta^2}\end{equation}
For the dark energy potential this is,
\begin{equation}\frac{S_E}{4} = 51.25\frac{{{M^2}}}{{{\delta ^2}}} = {\rm{1534.44}}\end{equation}
and the Universe will decay after $e^{S_E/4} = 2.505098\times 10^{666}$ Planck times (~$5.391\times (10^{-44} sec)$. Converting this to years, we get approximately $4.2823\times 10^{615}$ years. This is far greater than the age of the Universe which is $13.82\times 10^{9}$ billion years old. Thus, the dark energy potential is consistent with the stability of our Universe.

\subsection*{Quantum Computing for single radius Dark Energy Models }

In this section, we will be exploring the dark energy potential using IBM's open-source quantum computing software Qiskit. First, we will use classical methods to derive the Hamiltonian for the dark energy potential, and then apply quantum computing algorithms to solve for the ground state energy (the value for dark energy $\Lambda_4$). The main purpose of utilizing the quantum computational methods for our research is ultimately to help us understand the current limits and benefits of quantum computing.

The main procedure used for this exploration was the Variational Quantum Eigensolver (VQE). The VQE applies quantum mechanics' variational method to approximate the ground state energy of a system. This method repeatedly modifies an ansatz for a given wavefunction in an effort to get as accurate a result as possible. It attempts to create an upper bound for the ground state energy, giving an approximation for its actual value.

%\subsection*{The Hamiltonian for Dark Energy}

To begin, we derived the Hamiltonian for the dark energy potential and a matrix representation of the Hamiltonian using a classical computer (Mathematica software) and then imported them into the quantum computer code. To do so, we started by defining the number of qubits $n$, which establishes the size of the matrix for the Hamiltonian as $2^{n}\times 2^n$. For example, when we have 4 qubits, the matrix size will be $2^{4}\times 2^4$, or a $16\times 16$ matrix. We then created the matrix in the oscillator basis using Mathematica's SparseArray function. Lastly we applied the Variational Quantum Eigensolver  to the matrix and were able to compute the lowest eigenvalue. Below is the is the Hamiltonian used,
\begin{equation}H = \frac{p_{\phi}^2}{2} + V(\phi )\end{equation}
where $p_{\phi}$ is the momentum of the system. The Hamiltonian for the dark energy potential was determined to be,
%$$ H = \frac{p^2}{2} + \left[ \frac{7}{48(0.005)} \left( {e^\frac{-2\phi}{7.637626158259733}} -2{e^\frac{-\phi}{7.637626158259733}} + %{1.1845575279846237e^\frac{-2\phi}{7.637626158259733}} \right) \right]$$
\begin{equation}H = \frac{{\hat p_\phi ^2}}{2} + {e^{ - 4\hat \phi /a}}\left( {Q_4^2{e^{ - 8\hat \phi /a}} - k{e^{ - 2\hat \phi /a}} + {\Lambda _8}I} \right)\end{equation}
whwre $\hat p_{\phi}$ and $\hat \phi $ are represented by $2^n \times 2^n$ matrices.  We repeated the above process for 4, 5, and 6 qubits, thereby creating three separate sized matrices ($16\times 16$, $32\times32$, and $64\times64$, respectively) to test  the quantum computing methods. In the following, we compare the results of the classical computer $\Lambda_4$ value to the quantum computer's value.

%\subsection*{The Ground State}

In order to determine the ground state energy, $\Lambda_4$, of the above equation using the classical computer, we simply applied Mathematica's Eigenvalues function.
For the quantum computer, the procedure is more complex. It required importing the Hamiltonian into Python (the coding engine we used to run the quantum computing algorithms) and then creating the variational form using the Qiskit variational form EfficientSU2 (Special Unitary Group of degree 2). This is an essential part of the VQE algorithm in Qiskit. We used the "ry" and "rz" variational forms to allow for more accuracy in lieu of speed during runtime. The qubits were in full entanglement, meaning each qubit was entangled with all the other qubits with a CNOT gate. The depth of the circuit, or "reps", was set to 3, so that each variational form was repeated 3 times. As a visual of the variational form, we display the quantum circuit describing it below.
\begin{figure}[htp]
    \centering
    \includegraphics[width=13cm]{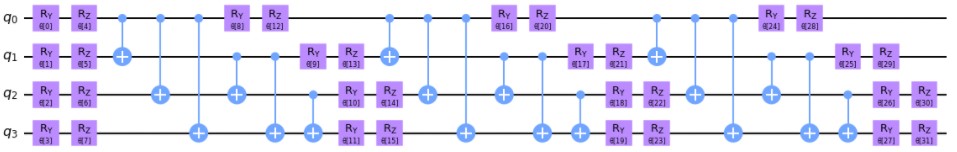}
    \caption{Image depicting the "ry" and "rz" variational form of an EfficientSU2 function applied to the 16x16 Hamiltonian for the dark energy potential. The qubits were in full entanglement using a CNOT gate and the repetitions were set to 3.}
    \label{fig:Variational Form}
\end{figure}

Next, we wrote a line of code to output the exact result for the ground state energy using NumPyEigensolver (see table 2). Then, we set up for the VQE code using the backend, "statevector\_simulator", together with the SLSQP optimizer (with maximum iterations at 600). To run the VQE code, we created an object which is an instance of the VQE class. We also created convergence plots to display an upperbound close to the exact value in figures 5 and 7. The results are recorded in table 2. Even with increased size of the matrix we were only able to get an upper bound of order $10^{-2}$ which is far above the actual value which was of order $10^{-6}$. The most likely reason for this is that the ansatz of the variational form does not have a strong overlap with the actual ground state wave function. It would be interesting to investigate other ansatz, as is doen in chemistry for example, to see if one can obtain a stronger overlap with the ground state wave function and a tighter bound on the ground state energy of the dark energy potential.

\begin{figure}[!htb]
\centering
\minipage{0.32\textwidth}
  \includegraphics[width=\linewidth]{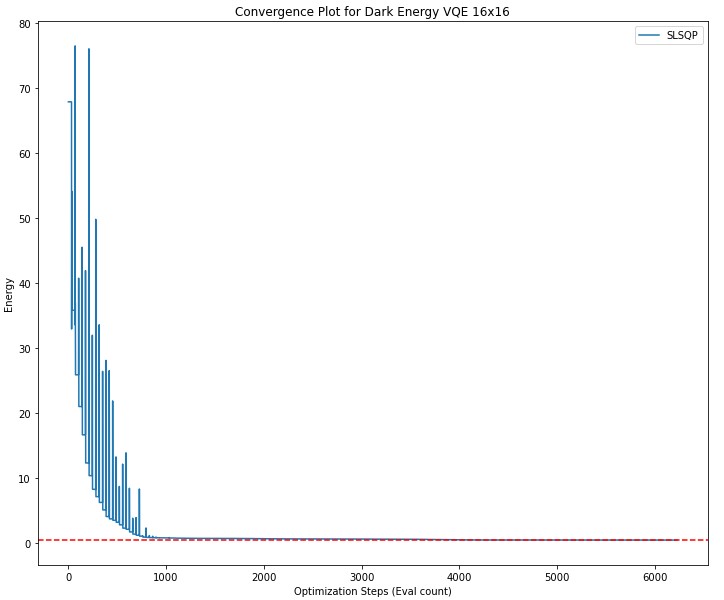}
%\label{fig:awesome_image1}
\endminipage\hfill
\minipage{0.32\textwidth}
  \includegraphics[width=\linewidth]{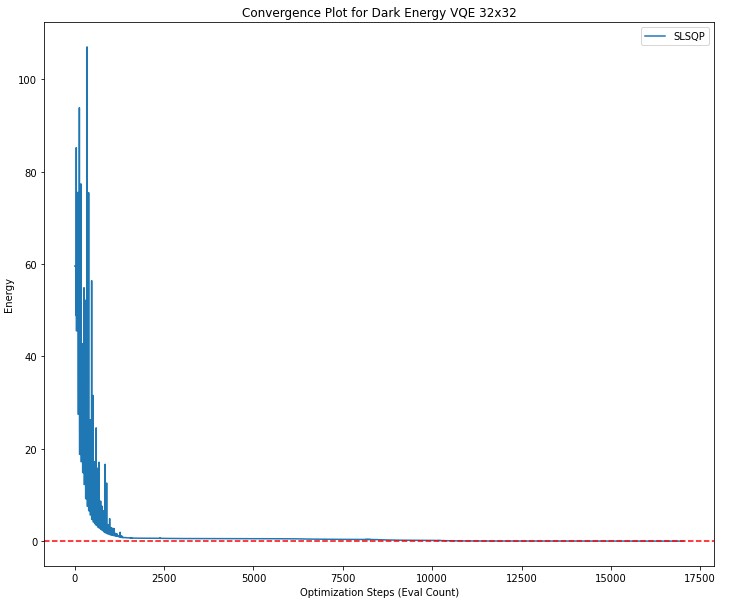}
\endminipage\hfill
\minipage{0.32\textwidth}%
  \includegraphics[width=\linewidth]{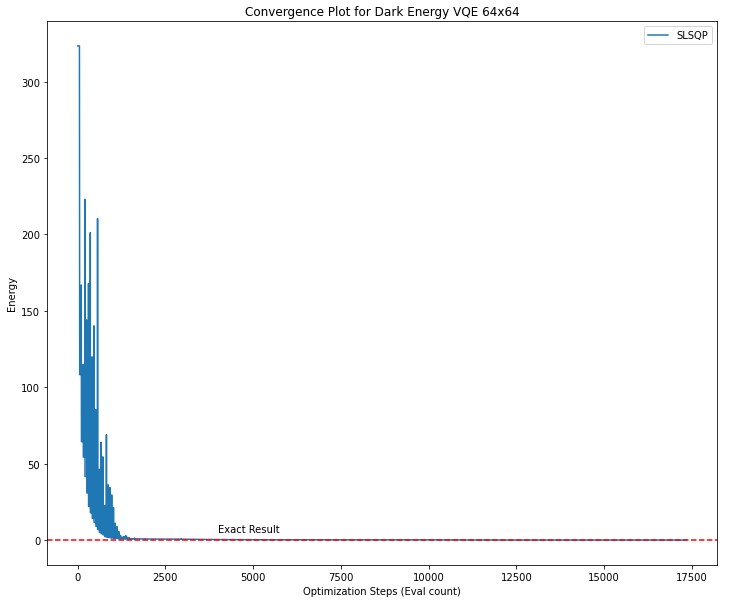}
 
\endminipage
\caption{(Left) Convergence plot for VQE calculation of dark energy potential with a single radius with four qubits or $16\times 16$ matrices. (Center) Convergence plot for VQE calculation of dark energy potential with a single radius with five qubits or $32\times 32$ matrices. (Right) Convergence plot for VQE calculation of dark energy potential with a single radius with six qubits or $64\times 64$ matrices.}
\end{figure}

%\begin{figure}[!htb]
%    \centering
%    \includegraphics[width=7.5cm]{Convergence_Plot_16.jpg}
%    \includegraphics[width=8cm]{Convergence_Plot_32.jpg}
 %   \includegraphics[width=8cm]{1Convergence_Plot_64.jpg}
 %   \caption{Convergence plot for the 16x16, 32x32, and 64x64 matrices of %VQE results on the quantum computer for $\Lambda_4$ in the dark energy %potential's Hamiltonian. In the upper right hand corner of each plot is a %legend containing the quantum computer's exact result for $\Lambda_4$, %the VQE result, and the program runtime.}
 %   \label{fig:Convergence}
%\end{figure}

%The table comparing the results of the quantum computer to the classical computer is shown in Figure 8. As suggested in the table, quantum computing still doesn't quite stand up to classical computing. Although it will, without a doubt, be an incredible method in the future, quantum computing hasn't quite surpassed classical computers just yet.

%\begin{figure}[htp]
%    \centering
 %   \includegraphics[width=14cm]{TableVQE.jpg}
%    \caption{Comparison of the classical computer (CC) results for the ground state energy, ($\Lambda_4$), of the Hamiltonian for dark energy potential and the quantum computer (QC) results (both exact and VQE). All units for $\Lambda_4$ values are in $M_{Planck}^{4}$. Recall the actual known value is $\Lambda_4 \approx 10^{-120} M_{Planck}^{4}$.}
 %   \label{fig:Table VQE}
%\end{figure}
\begin{table}
\begin{center}
\begin{tabular}{ |c|c|c|c|c|c| } 
 \hline
Matrix Size & Qubits &   Exact Ground State Energy & VQE Result  \\
\hline
 $16\times 16$ & 4 & .43791588 & .43794824 \\ 
 \hline
 $32\times 32$ & 5 & .00285585 & .00681012 \\ 
 \hline
 $64\times 64$ & 6 & .00000112 & .00916398 \\ 
\hline
\end{tabular}
\end{center}
\caption{Results from the VQE computation for the Dark energy single radius model using the SLSQP optimizer and  4, 5 and 6  qubits.}
\end{table}

\newpage
\subsection*{Two radius dark energy potential}

For an internal space $S^2 \times S^2$ the dark energy potential is written as:
\begin{equation}V({R_1},{R_2}) = \mu _1^4R_1^{-2}R_{2}^{-2}\left( {\frac{{Q_1^2}}{{R_1^4}} + \frac{{Q_1^2}}{{R_1^4}} - \frac{1}{{R_1^2}} - \frac{1}{{R_2^2}} + {\Lambda _8}} \right)\end{equation}
The first factor comes from redefining the metric so the Einstein-Hilbert action has no extra dimensional radii multiplying it, the first term in brackets comes from the two independent fluxes through the two $S^2$'s, the second term comes from their curvature and the last term comes from an eight dimensional cosmological constant.

Defining two scalar fields as in \cite{Brown:2014sba} from:
\[{\mu _2}\log ({R_1}) = \frac{1}{{\sqrt 2 }}{\phi _1} - \frac{1}{{\sqrt 2 }}\frac{{\sqrt 3  - 1}}{{2\sqrt 3 }}({\phi _1} + {\phi _2})\]
\begin{equation}{\mu _2}\log ({R_2}) = \frac{1}{{\sqrt 2 }}{\phi _2} - \frac{1}{{\sqrt 2 }}\frac{{\sqrt 3  - 1}}{{2\sqrt 3 }}({\phi _1} + {\phi _2})\end{equation}
The potential is written as:
\[V({\phi _1},{\phi _2}) = \mu _1^4 e^{ \left[ { - 2\left( {\frac{1}{{\sqrt 2 }}{\phi _1} - \frac{1}{{\sqrt 2 }}\frac{{\sqrt 3  - 1}}{{2\sqrt 3 }}({\phi _1} + {\phi _2})} \right)/{\mu _2}} \right]}e^{ \left[ { - 2\left( {\frac{1}{{\sqrt 2 }}{\phi _2} - \frac{1}{{\sqrt 2 }}\frac{{\sqrt 3  - 1}}{{2\sqrt 3 }}({\phi _1} + {\phi _2})} \right)/{\mu _2}} \right]} \cdot \]
\[\left( {\left( {Q_1^2e^ {\left[ { - 4\left( {\frac{1}{{\sqrt 2 }}{\phi _1} - \frac{1}{{\sqrt 2 }}\frac{{\sqrt 3  - 1}}{{2\sqrt 3 }}({\phi _1} + {\phi _2})} \right)/{\mu _2}} \right]} + Q_2^2e^{ \left[ { - 4\left( {\frac{1}{{\sqrt 2 }}{\phi _1} - \frac{1}{{\sqrt 2 }}\frac{{\sqrt 3  - 1}}{{2\sqrt 3 }}({\phi _1} + {\phi _2})} \right)/{\mu _2}} \right]}} \right) + } \right.\]
\begin{equation}\left. {\left( { - e^ {\left[ { - 2\left( {\frac{1}{{\sqrt 2 }}{\phi _1} - \frac{1}{{\sqrt 2 }}\frac{{\sqrt 3  - 1}}{{2\sqrt 3 }}({\phi _1} + {\phi _2})} \right)/{\mu _2}} \right]} - e^{ \left[ { - 2\left( {\frac{1}{{\sqrt 2 }}{\phi _1} - \frac{1}{{\sqrt 2 }}\frac{{\sqrt 3  - 1}}{{2\sqrt 3 }}({\phi _1} + {\phi _2})} \right)/{\mu _2}} \right]} + {\Lambda _8}} \right)} \right)\end{equation}
The potential is plotted in figure 8.
\begin{figure}
\centering
\minipage{0.5\textwidth}
  \includegraphics[width=\linewidth]{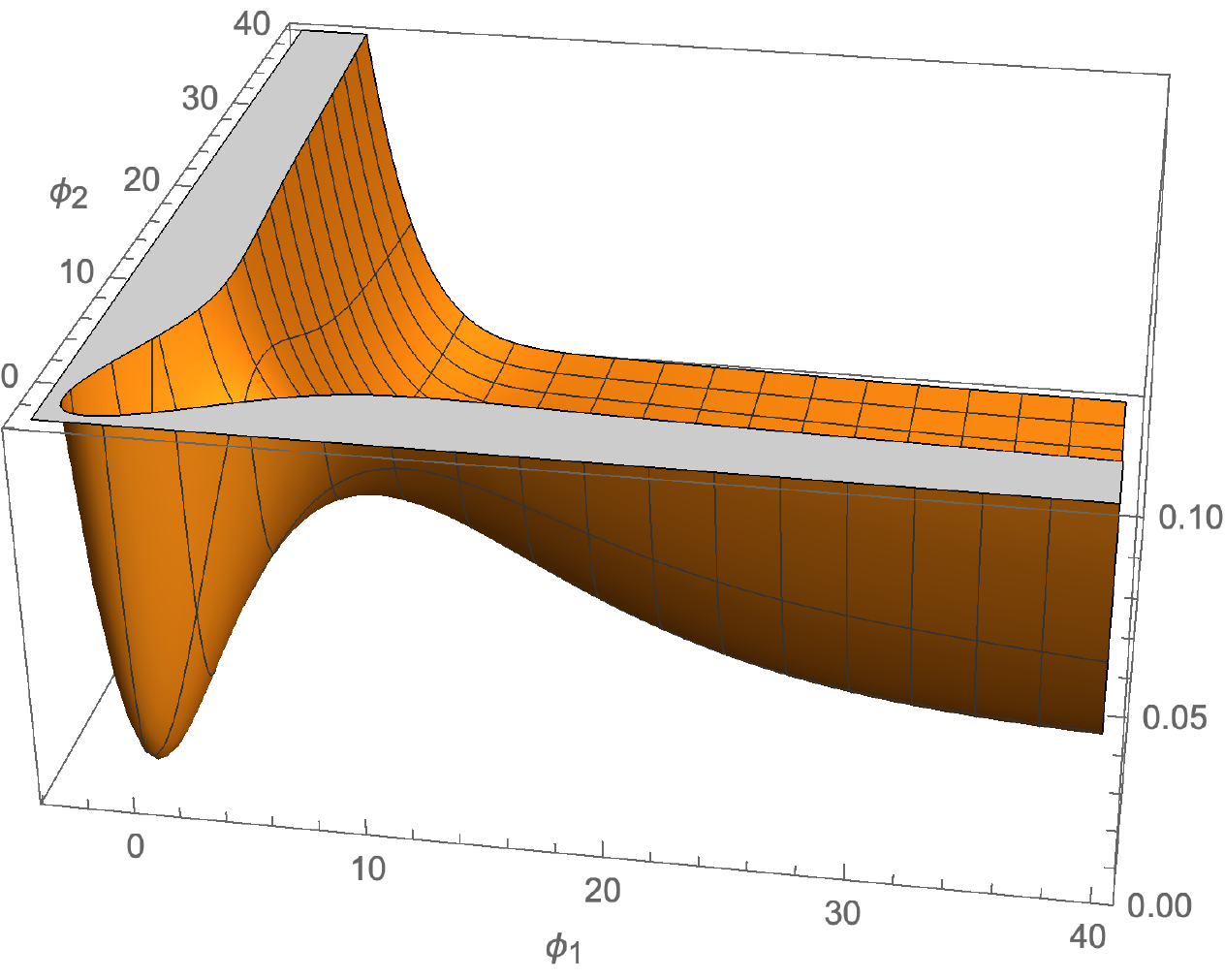}
%\label{fig:awesome_image1}
\endminipage\hfill
\minipage{0.5\textwidth}
  \includegraphics[width=\linewidth]{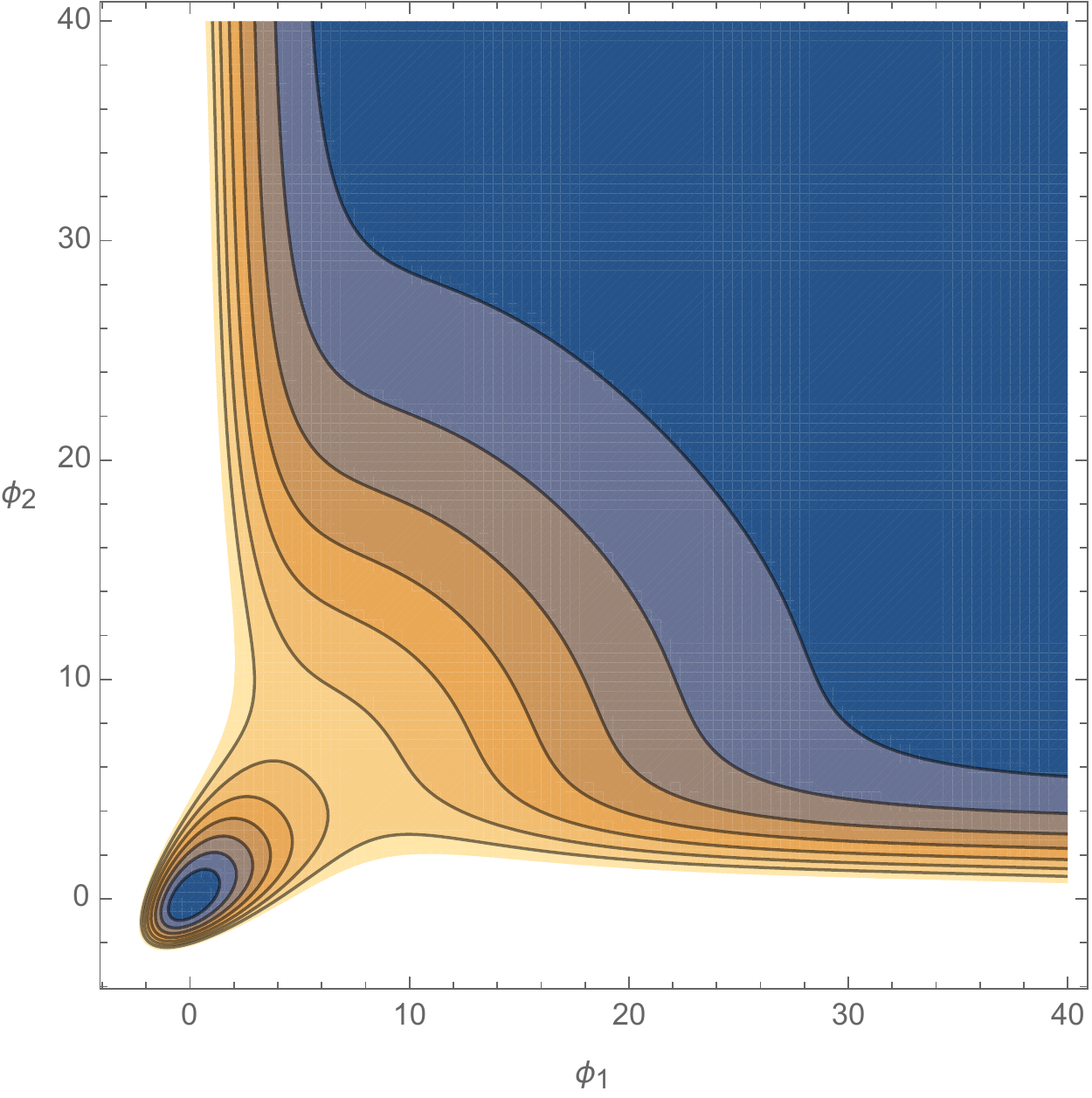}
\endminipage\hfill
\caption{(Left) 3D plot of a form of the Dark Energy potential for internal space $S^2 \times S^2$.  (Right) Contour plot of a form of the Dark Energy  potential for internal space $S^3 \times S^2$.}
\end{figure}

\subsection*{Quantum computing for two radius dark energy potential}

In figure 9 and table 3 we show the results of the VQE computation for the dark energy model with two radii and parameters $\mu_1^4 = 200$ and $\mu_2 =8$ in Planck units.. This  model requires a tensor product Hilbert space associated with the variables $\phi_1, \phi_2$. If we use 4 qubits for each variable the product Hilbert space will use 8 qubits and the Hamiltonian oberator will be realized using the oscillator basis as a $256 \times 256$ matrix. The two radius dark energy model requires far more Pault terms to represent the model in terms of qubits than the single radius model. The upper bound on the vacuum energy was of order $10^{-1}$ versus $10^{-5}$ for the exact value. Again this was most likely because the variational ansatz does not have a strong enough overlap with the true ground state.
\begin{figure}[h!]
\centering
  \includegraphics[width=.4\textwidth]{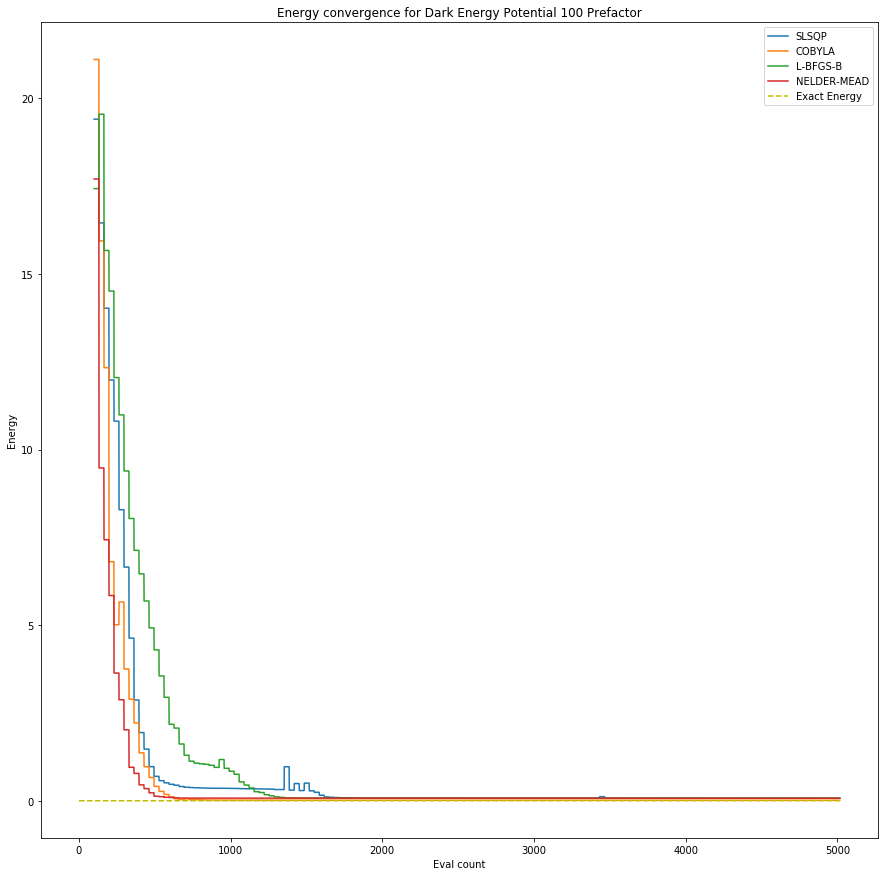}
  \caption{Convergence plots of the ground state energy for Dark Energy two radius model  using  8 qubits. }
\end{figure}
\begin{table}
\begin{center}
\begin{tabular}{ |c|c|c|c|c|c| } 
 \hline
Matrix Size & Qubits & Pauli Terms & Exact Ground State Energy & VQE Result  \\
\hline
 $256 \times 256$ & 8 & 15115 & 0.00004821 & 0.07171467\\ 
 \hline
\end{tabular}
\end{center}
\caption{Results from the VQE computation for the Dark energy two radius model  using the NELDER-MEAD optimizer and  8 qubits. The energy converges to 0.07171467 in Planck units.}
\end{table}

\newpage
\section{Dark Matter Cosmology}

Why do scientists believe that the universe is composed of some unknown matter that has never been observed? And more so, how do we know that this form of matter comprises more of the universe than ordinary matter? Theorized by Fritz Zwicky in 1933 to explain why galaxy clusters were not flying apart, dark matter's first direct evidence came in the 1970's with Vera Rubin and Kent Ford's observation of gravitational interactions in galaxy rotation curves. Rubin and Ford witnessed a surprising sight, the stars rotating at the edges of Andromeda galaxy had similar velocities to those at the center, seemingly violating Newtonian physics. This could easily be explained if there was some invisible matter contributing to the gravitational force. Similarly, galaxies in galaxy clusters have been found to have higher velocity than expected. Again the simplest solution is that there is some form of matter that we can not detect.

Dark matter's existence is further evidenced in its imprint on the cosmic microwave background (CMB). The early universe was a hot dense plasma and dark matter flowed to regions of increased matter density drawn by gravity. For ordinary baryonic matter, the push of radiation pressure fought against the pull of gravity leading to oscillations dependent on the density of matter in that region. Dark matter did not build up radiation pressure and therefore clumped more efficiently. This can help explain the large scale structure of our universe today. At some point, the universe cooled enough that photons could decouple from ordinary matter and it is this moment captured in the CMB. Specifically, the universe's composition is encoded in the CMB's power spectrum. The first peak corresponds to the total energy in the universe, a combination of dark energy, dark matter and baryons. The size of the second peak relative to the first gives information about the baryon content of the universe, while the subsequent peaks show the relative amount of dark matter.

\subsection*{Self-interacting dark matter}
Cold dark matter (CDM) models have received several decades of attention but simulations based on these models break down at explaining the distribution of dark matter in small galaxies. For instance, an issue with CDM is that simulations predict centers of galaxies and clusters with overly dense cores and too many halos within dwarf galaxy groups as compared to observation. Yet, CDM models predict large scale structure to a remarkably accurate degree. The issue with CDM can be resolved by turning our attention to self-interacting dark matter (SIDM) \cite{Spergel:1999mh}
\cite{Forestell:2016qhc}
\cite{Faraggi:2000pv}. SIDM simulations alleviate the core cusp problem as they affect structure $\leq$ 1 Mpc while still maintaining the accuracy of the CDM model at large distances. Self-interacting dark matter interacts with itself potentially through gravity, with a large scattering cross-section but negligible annihilation or dissipation. There are exacting constraints on the interactions between dark and ordinary matter, as well as on long range forces between dark matter particles. However, if the dark matter annihilation cross-section is much smaller than the scattering cross-section there are relatively few constraints on short-range dark matter self-interactions.  

 In this paper, we will primarily focus on SU(2) glueball dark matter candidates from a gauge-theory non-abelian hidden sector. Ordinary visible sector glueballs are predicted as bound states by Quantum Chromodynamics (QCD) and can be expected for hidden non-Abelian gauge forces for self interacting dark matter. They are a theorized particle type  due to the emergence of composite particles from the strong interactions between dark gluons, the analog of the force mediators of the strong nuclear force. The lightest hidden sector glueball states are strongly interacting among themselves but very weakly interacting with the Standard Model states and have no lighter state to decay into. Spergel and Steinhardt \cite{Spergel:1999mh} also make a number of predictions for the properties of galaxies in a self-interacting dark matter cosmology: (1) the centers of halos are spherical; (2) dark matter halos will have cores; and (3) there are few dwarf galaxies in groups but dwarfs persist in lower density environments; and, (4) the halos of dwarf galaxies and galaxy halos in clusters will have radii smaller than the gravitational tidal radius which are all consistent with observation. For these reasons, glueballs are a good self-interacting dark matter candidate to explore. Another self interacting model we will explore is conformally coupled scalar fields which can also be self interacting through a quartic interaction and are somewhat simpler than dark glueballs \cite{Kapetanakis}\cite{Cavaglia:1996ek}\cite{Maleknejad:2011jw}.
 
%\section{Models of self interacting dark matter}
\subsection*{Model one: conformally coupled scalar fields}

In this model the visible sector consists of a conformally coupled scalar field $X$ with self interactions and the dark sector consists of another conformally coupled  scalar field $Y$ with self interactions. A benefit of this model is that scalar fields do not affect the the homogeneity and isotropy required of cosmological models. We shall include a small coupling term between the the visible and dark sectors which will be eighth order in the fields. The Lagrangian for this theory is given by:
$L = L_x + L_Y+L_{mix}$
where:
\[{L_X} = \sqrt { - g} \left( { - \frac{1}{2}{g^{\mu \nu }}{\partial _\mu }X{\partial _\nu }X - \frac{1}{{12}}R{X^2} - {\lambda _X}{X^4}} \right)\]
\[{L_Y} = \sqrt { - g} \left( { - \frac{1}{2}{g^{\mu \nu }}{\partial _\mu }Y{\partial _\nu }X - \frac{1}{{12}}R{Y^2} - {\lambda _Y}{Y^4}} \right)\]
\begin{equation}{L_{mix}} = \sqrt { - g} \left( { - {\lambda _{mix}}{X^4}{Y^4}} \right)\end{equation}
now using the ansatz:
\begin{equation}d{s^2} = {a^2}(t)( - d{t^2} + d\Omega _3^2)\end{equation}
where $d\Omega_3^2$ is the volume element for a unit three sphere.
After rescaling $X$ and $Y$ by $1/a(t)$ the Hamiltonian for the matter dark matter system is given by:
\begin{equation}H = \frac{{P_X^2}}{2} + \frac{{{X^2}}}{2} + \frac{{P_Y^2}}{2} + \frac{{{Y^2}}}{2} + {\lambda _X}{X^4} + {\lambda _Y}{Y^4} + \frac{1}{{{a^4}}}{\lambda _{mix}}{X^4}{Y^4}\end{equation}
This is the form of the Hamiltonian we will use for the quantum computer simulation. 

\subsection*{Model two: SU(2) gauge fields }
For the second model we will consider the visible sector given by a $SU(2)$ gauge field and the dark matter given by another $SU(2)$ gauge field. 
We will also include a theta term in the dark sector as this term is not constrained like the the theta term in the visible sector for strong CP violation. The Lagrangian for this theory is given by:
$L=L_X+L_Y+L_{mix}+L_{theta}$
where:
\[{L_X} = \sqrt { - g} \left( { - \frac{1}{4}F_{\mu \nu }^{Xa}F_a^{X\mu \nu }} \right)\]
\[{L_Y} = \sqrt { - g} \left( { - \frac{1}{4}F_{\mu \nu }^{Ya}F_a^{Y\mu \nu }} \right)\]
\[{L_{mix}} =  - \sqrt { - g} \frac{{{\lambda _{mix}}}}{{384}}\left( {\frac{1}{4}F_{\mu \nu }^{Ya}F_a^{Y\mu \nu }} \right)\left( {\frac{1}{4}F_{\mu '\nu '}^{Ya'}F_{a'}^{Y\mu '\nu '}} \right)\]
\begin{equation}{L_{theta}} =  - {\theta _Y}\sqrt { - g} {\varepsilon ^{\mu \nu \lambda \sigma }}F_{\mu \nu }^{Ya}F_{\lambda \sigma }^{Ya}\end{equation}
Now using the ansatz for the gauge fields:
\[F_{0i}^{Xa} = \dot X\delta _i^a,F_{ij}^a =  - {X^2}\varepsilon _{ij}^a\]
\begin{equation}F_{0i}^{Ya} = \dot Y\delta _i^a,F_{ij}^a =  - {Y^2}\varepsilon _{ij}^a\end{equation}
The Lagrangian can be written as:
\begin{equation}L = \frac{1}{2}{{\dot X}^2} - g_X^2{X^4} + \frac{1}{2}{{\dot Y}^2} - g_Y^2{Y^4} - \frac{{{\lambda _{mix}}}}{{{a^4}}}{\left( {\dot X + {X^2}} \right)^2}{\left( {\dot Y + {Y^2}} \right)^2} - {\theta _Y}\dot Y{Y^2}\end{equation}
The Hamiltonian is then given by:
\begin{equation}H = {P_X}\dot X + {P_Y}\dot Y - L\end{equation}
where
\[{P_X} = \dot X - 2\frac{{{\lambda _{mix}}}}{{{a^4}}}\left( {\dot X + {X^2}} \right){\left( {\dot Y + {Y^2}} \right)^2}\]
\begin{equation}{P_Y} = \dot Y - 2\frac{{{\lambda _{mix}}}}{{{a^4}}}{\left( {\dot X + {X^2}} \right)^2}\left( {\dot Y + {Y^2}} \right) - {\theta _Y}{Y^2}\end{equation}
Now $\lambda_{mix}$ is considered small so we can approximate:
\[{P_X} = \dot X\]
\begin{equation}{P_Y} = \dot Y - {\theta _Y}{Y^2}\end{equation}
The Hamiltonian is then
\begin{equation}H = \frac{1}{2}{P_X}^2 + g_X^2{X^4} + \frac{1}{2}{\left( {{P_Y} + \theta {Y^2}} \right)^2} + g_Y^2{Y^4} + \frac{{{\lambda _{mix}}}}{{{a^4}}}{\left( {{P_X} + {X^2}} \right)^2}{\left( {{P_Y} + {\theta _Y}{Y^2} + {Y^2}} \right)^2} + {\theta _Y}\left( {{P_Y} + {\theta _Y}{Y^2}} \right){Y^2}\end{equation}
The form of the Hamiltonian is more complicated than the model with two conformal scalar fields. We set the theta angle in the dark sector to zero in this paper but will analyze the effects of this term in future work.

\subsection*{Quantum computing for dark matter models}
For calculations, we used a mix of classical and quantum computing. Previous application of quantum computing to dark matter simulation can be found in \cite{Mocz:2021ehj}.  In this section modeling for the two SIDM frameworks was done with a mix of classical and quantum analysis, the methodology of which is described below.

%\subsection{Classical computing for SIDM models }
We derived Hamiltonians for both models using the Lagrangian above. We determined the exact energy levels and their associated wave functions by solving differential equations of the systems in Mathematica. The exact ground state energy calculated for a continuous system was used to compare results with that derived from quantum computing.

%\subsubsection{Creation of Hamiltonian mappings}
Notably, to solve for the ground state energy using quantum computing one needs to create a Hamiltonian mapping (we used Mathematica) to import into Qiskit. As the Hilbert space of a quantum field theory is infinite-dimensional, we needed to simplify by discretizing space. This was accomplished by creating matrix representations of Hilbert space operators in an oscillator basis. The position operator and momentum operators were created in a way that mimics the ladder operators of quantum mechanics. After creating a matrix $\hat{a}$, one can create position and momentum in the manner below:

   \[ \hat{Q} = \frac{1}{\sqrt{2}}(a + a^{\dagger}) \]
    \begin{equation}
    \hat{P} = \frac{i}{\sqrt{2}}(-a + a^{\dagger}) 
\end{equation}
where $a$ and $a^{\dagger}$ are the harmonic oscillator ladder functions. For both models of SIDM, we used bosonic operators to simulate the system (which can also be thought of as a two-dimensional  single particle system). Bosons can exist in an infinite number of excited states, and thus their matrix operators can be extended indefinitely. To simulate two bosonic fields, I used tensor products, namely the Kronecker product to create an $N\times N$ matrix for momentum and position: 
\[
    \hat{X} = \hat{Q} \otimes \hat{I}, \quad \hat{Y} = \hat{I} \otimes \hat{Q} \]
 \begin{equation}   \hat{P_X} = \hat{P} \otimes \hat{I}, \quad \hat{P_Y} = \hat{I} \otimes \hat{P}
\end{equation}

%\subsection{Quantum computing for SIDM models}
%In 1982, Richard Feynman posited the core idea for quantum computing. Stating, "Now it turns out, as far as I can tell, that you can simulate this with a quantum system, with quantum computer elements (...) If we disregard the continuity of space and make it discrete, and so on, as an approximation (the same way as we allowed ourselves in the classical case), it does seem to be true that all the various field theories have the same kind of behavior, and can be simulated in every way, apparently, with little latticeworks of spins and other things"[14]. Where classical computers rely on bits (binary digits) which are always either 0 or 1,  quantum computers rely on qubits which exist as 0 or 1 but also as superpositions of the two states. Superposition and entanglement of the qubits, along with the usage of quantum circuits to manipulate the state of the qubit, can possibly allow for more advanced computing capabilities. However, computations can be limited as the number of states is exponential to the size of the system.

For quantum computing results, we used IBM's Quantum information science kit (Qiskit), an open-source software stack with Python interface that communicates with an IBM quantum computer in real time. All analysis was done with a statevector simulator. The statevector simulator executes an ansatz of a Qiskit quantum circuit and then returns the final quantum statevector of the simulation. We  used the sequential least squares programming (SLSQP) optimizer and NELDER-MEAD optimizer which returned accurate results.
\begin{figure}[ht]
    \centering
    \includegraphics[scale=.22]{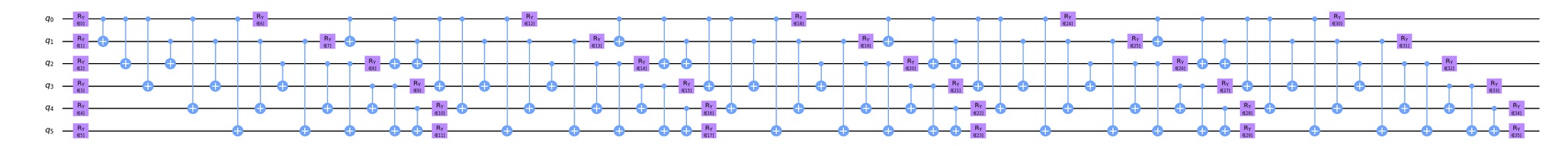}
    \caption{Quantum circuit generated for ansatz of Model one using VQE algorithm. 6 qubits.}
    \label{fig:QC6}
\end{figure}

In Fig 10, one can see the quantum gates applied to the initial state. We utilized Pauli and SU2 operators as well as an Ry gate, which is a rotational operator. The algorithm employed full entanglement with Controlled NOT (CNOT) gates. The CNOT gate is a two-qubit operation. The first qubit is the control qubit and the second qubit is the target qubit. The CNOT gate flips the target qubit only if the control qubit is $|1 \rangle $.  Finally, the depth for the variational form was set to 5 repetitions.

%\subsubsection{VQE algorithm}
The main algorithm used was the variational quantum eigensolver (VQE), which is a hybrid classical-quantum algorithm that variationally determines the ground state energy of a Hamiltonian. The expectation value of the energy is computed via a quantum algorithm, while the energy is minimized with a classical optimization algorithm. For this paper I mainly used the VQE to find the ground state energy convergence. First, I created a Hamiltonian in a computational oscillator basis as described above. Then, the VQE applies a variational ansatz. Figure 10 is the representation of this ansatz as a quantum circuit. Given the circuit, the VQE algorithm measures the expectation value of the Hamiltonian and varies the circuit parameters until the energy is minimized. 

%\section{Self-Interacting Dark Matter Results}

%\begin{figure}[ht]
 %   \centering
  %  \subfigure{\includegraphics[scale = 0.22]{DMmodel1_6q.jpg}}
 %   \subfigure{\includegraphics[scale = 0.22]{DMmodel1_8q.jpg}}
 %   \caption{Convergence of eigenvalue to $E_0$ (ground state energy) for Dark Matter model one using SLSQP optimizer and 8 qubits. Energy converges to 1.01205913.}
 %   \label{fig:Model1}
%\end{figure}
%\begin{figure}[ht]
 %   \centering

  %  \includegraphics[width=.5\textwidth]{DMmodel1_8q.jpg}
 %   \caption{Convergence of eigenvalue  $E_0$ (ground state energy) for Dark Matter model one using SLSQP optimizer and 8 qubits. }
 %   \label{fig:Model1}
%\end{figure}

%\subsection{SIDM: Model One}
Using Mathematica, we solved for the eigenvalues (energy) and eigenfunctions (wavefunctions) for a $V(X,Y) = \frac{1}{2}(X^2 + Y^2) + \lambda_X X^4 + \lambda_Y Y^4 + \frac{1}{{{a^4}}}{\lambda _{mix}}{X^4}{Y^4}$   potential and the first four eigenfunctions are shown in figure 11.
\begin{figure}[!htb]
\centering
\minipage{0.5\textwidth}
  \includegraphics[width=\linewidth]{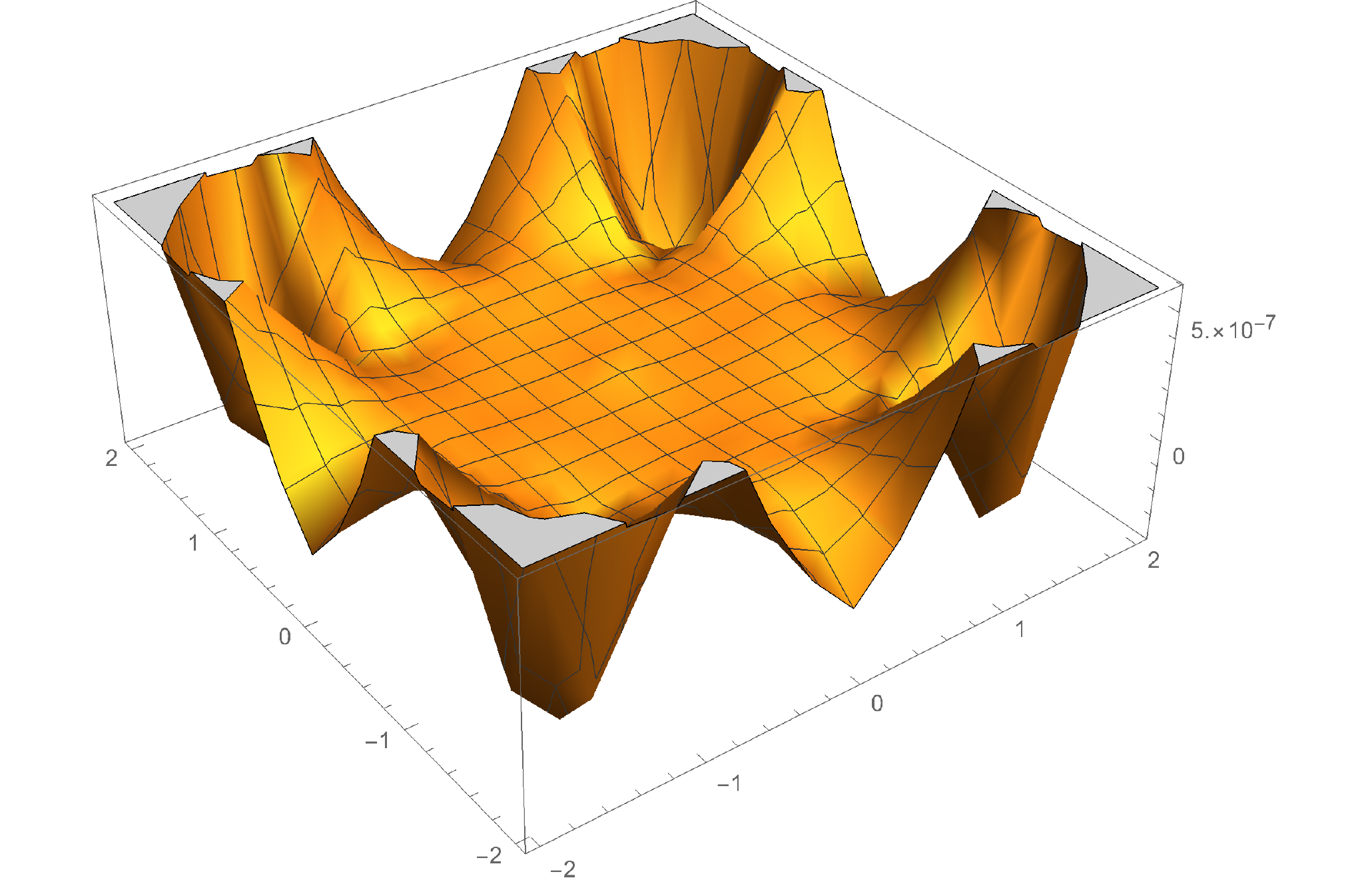}
%\label{fig:awesome_image1}
\endminipage\hfill
\minipage{0.5\textwidth}
  \includegraphics[width=\linewidth]{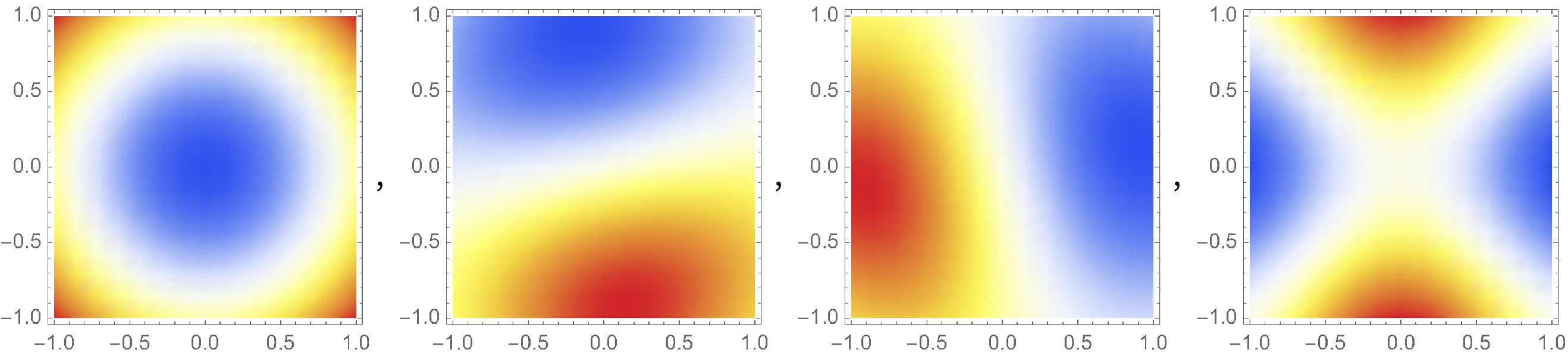}
\endminipage\hfill
\caption{(left) Plot of potential for Model one (right) First four eigenfunctions for Model one.}
\end{figure}
The Hamiltonian for Model one is given by:
\begin{equation}H = \frac{{P_X^2}}{2} + \frac{{{X^2}}}{2} + \frac{{P_Y^2}}{2} + \frac{{{Y^2}}}{2} + {\lambda _X}{X^4} + {\lambda _Y}{Y^4} + \frac{1}{{{a^4}}}{\lambda _{mix}}{X^4}{Y^4}\end{equation} 
The matrix representation is based on the Hamiltonian of the system (see above), with $a = 1$; $\lambda_X = \lambda_Y = 0.005; \lambda_{mix} = 0.001 $ for calculations. 

%Notice that the Hamiltonian has $X^4Y^4$ mixing term in the potential. It is interesting to see the change in wavefunction with additional terms terms. Fig 6A shows ground state wavefunction with $X^2+Y^2$, the wave function is perturbed with the addition of $X^4 + Y^4$ (Fig 6B), Fig 6C introduces a Pullen-Edmonds potential term [16] of $X^2Y^2$, wh  ile the last image (Fig 6D) shows the wavefunction associated with Model 1. A note of caution here, I increased the coupling to a much higher value of 0.1 in order to better see the differences. We should however use a lower coupling constant as dark matter and baryonic matter could only have minimal interaction.

%\begin{figure}
% \centering
 %   \subfigure{\includegraphics[scale = 0.18]{1.pdf}}
 %   \subfigure{\includegraphics[scale = 0.18]{2.pdf}}
 %   \subfigure{\includegraphics[scale = 0.18]{3.pdf}}
 %   \subfigure{\includegraphics[scale = 0.18]{4.pdf}}
 %   \caption{ Evolution of wavefunction with the addition of terms. (6A)$X^2+Y^2$ (6B) Addition of $X^4 + Y^4$ self interaction terms. (6C) Addition of $X^2Y^2$ coupling term (6D) Removal of second order coupling term, substituting $X^4Y^4$} 
  %  \label{fig:Wave}
%\end{figure}

Table 4 shows the results of VQE analysis for Dark matter model one. It provides a comparison of the exact values for ground state energy (column 6) and VQE values (column 5). The terms in parenthesis in column 6 are the exact values calculated by Python for the discrete Hamiltonian matrices, while the values from Mathematica were calculated for a continuous system with a differential equation. Note the calculated percent error (column 7) in order to understand the accuracy of the VQE results as well as any possible influence of matrix size on efficiency. As one can see the results found were fairly accurate. We found similar results for model two in table 5. The final variational wave function in quantum circuit for is displayed in figure 10. Convergence graphs for various optimizers for model one and two are show in figure 12. 
\begin{table}[]
    \centering
    \begin{tabular}{|p{2cm}|p{1cm}|p{1cm}|p{1cm}|p{2cm}|p{2cm}|p{1cm}|}
    \hline
    \multicolumn{7}{|c|}{Model One: VQE results}\\
    \hline
      Matrix Size  & Qubits & Pauli Terms & Time (s) & VQE & Exact & Diff ($\%$) \\
      \hline
      16x16  & 4 & 25 & 0.108 & 1.01208469 & 1.01206 (1.01208468) & 0.0020\\
        \hline
       64x64  & 6 & 361 & 7.958 & 1.01220476 & 1.01206 (1.01205876) & 0.0138 \\
         \hline
         256 x 256  & 8 & 3025 & 54.900 & 1.012124134 & 1.01206 (1.01205913) & 0.0059\\
           \hline
    \end{tabular}
    \caption{VQE results for Model One using SLSQP optimizer. In Ground State Energy Exact column, first value is exact value from Mathematica while second value is value given by Python from discretized Hamiltonian input. }
    \label{tab:VQEM1}
\end{table}
\begin{table}
\begin{center}
\begin{tabular}{ |c|c|c|c|c|c| } 
 \hline
Model & Qubits & Pauli Terms & Exact Ground State Energy & VQE Result  \\
\hline
Daark matter model two  & 8 & 3024 & 1.01500000 & 1.01500466\\ 
 \hline
\end{tabular}
\end{center}
\caption{VQE results for Model two using SLSQP optimizer.}
\end{table}
%\begin{figure}[h!]
%\centering
%  \includegraphics[width=\textwidth]{Circuitforall4Models.png}
%  \caption{Quantum circuit that represents the variational wave form for  dark matter model two and 8 qubits.}
%\end{figure}
\begin{figure}[!htb]
\centering
\minipage{0.5\textwidth}
  \includegraphics[width=\linewidth]{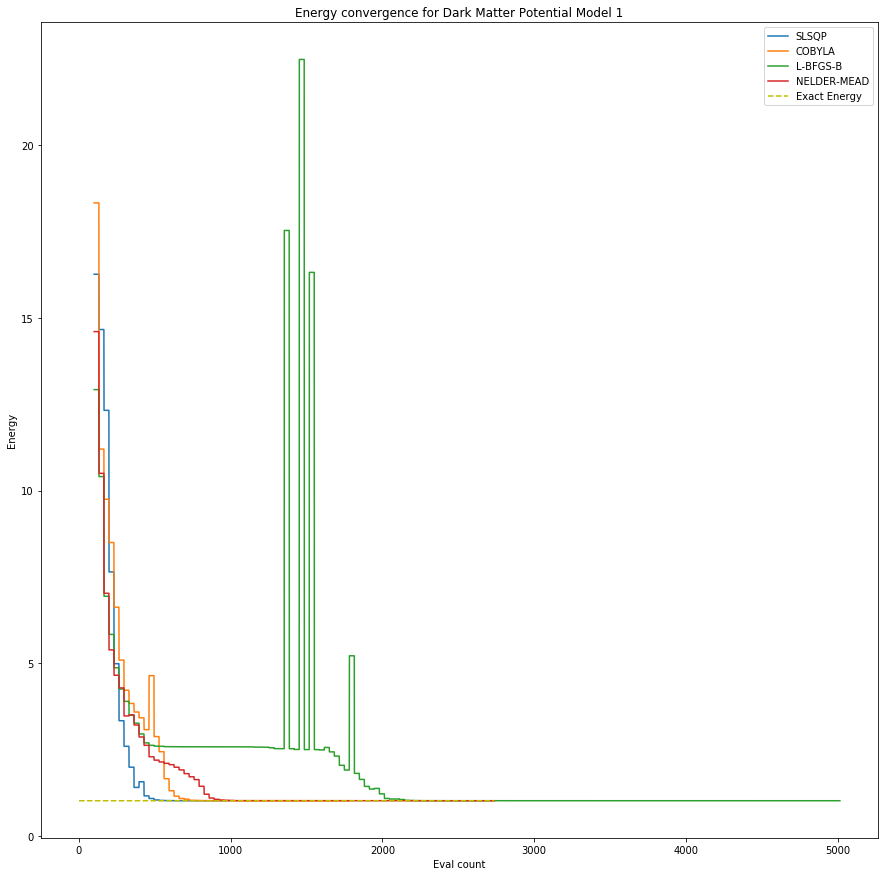}
%\label{fig:awesome_image1}
\endminipage\hfill
\minipage{0.5\textwidth}
  \includegraphics[width=\linewidth]{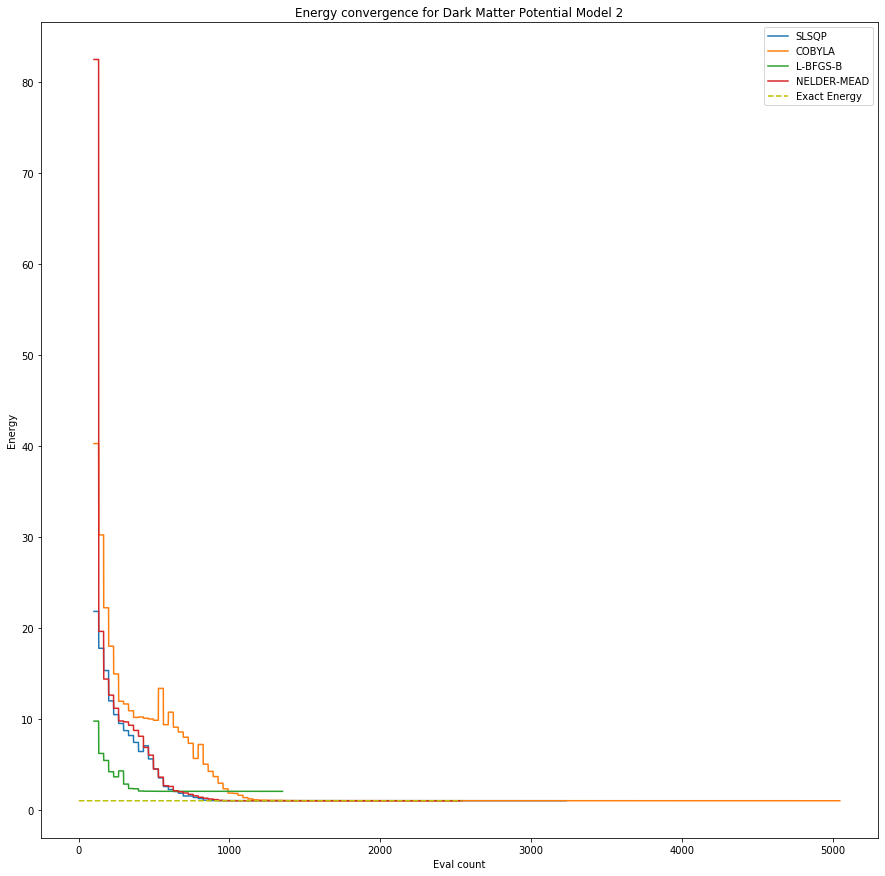}
\endminipage\hfill
\caption{Convergence plots of the ground state energy for Dark Matter  Model one (left) and Model two (right) using various  optimizers and 8 qubits. The SLQP optimizer converges rapidly to an upper bound close to the exact value}
\end{figure}
%\begin{figure}[h!]
%\centering
%  \includegraphics[width=.5\textwidth]{100PrefactorConvergence.png}
%  \caption{Convergence plots of the ground state energy for Dark Energy 100 Prefactor using the NELDER-MEAD optimizer. 8 qubits. The energy converges to 0.000160542.}
%\end{figure}

%\subsection{SIDM: Model Two}
%Again I used the matrix representation based on the Hamiltonian of the system: 
%\[H = \frac{1}{2}{P_X}^2 + g_X^2{X^4} + \frac{1}{2}{\left( {{P_Y} + \theta_Y {Y^2}} \right)^2} + g_Y^2{Y^4} + \frac{{{\lambda _{mix}}}}{{{a^4}}}{\left( {{P_X} + %{X^2}} \right)^2}{\left( {{P_Y} + {\theta _Y}{Y^2} + {Y^2}} \right)^2} + {\theta _Y}\left( {{P_Y} + {\theta _Y}{Y^2}} \right){Y^2}\] 
%As $\hat{P}_{X,Y}$ do not commute with X or Y, care needs to be taken when expanding out the Hamiltonian to ensure that the matrices created are Hermitian. This %created some problems for my analysis. Further work should be done to explore this model. 

\newpage
\section{Quantum Cosmology on a quantum computer}

Quantum computing maps a quantum mechanical computation to a physical system that is in a quantum regime. The physical system thus can perform a Lorentzian path integral and because of the mapping, the Lorentzian path integral of that system can also be computed on a quantum computer. This has potential advantages for quantum cosmology because Lorentzian path integrals do not depend on integration contours like Euclidean path integrals. Also the Euclidean action for quantum cosmology is unbounded from below which makes Monte Carlo calculations on classical computers problematic. Thus Lorentzian path integrals on quantum computers may have an advantage in quantum cosmology.

\subsection*{ Quantum cosmology with a scalar field }

One of the simplest and most important cosmological models is four dimensional gravity coupled to a scalar field which has implications for inflationary cosmology. The Einstein-Hilbert action with a free massless scalar field is given by:
\begin{equation}S = \int {{d^4}x\sqrt { - g} } \left[ {\frac{1}{{16\pi G}}\left( {R - 2\Lambda } \right) - \frac{1}{2}{g^{\mu \nu }}{\partial _\mu }\phi {\partial _\nu }\phi } \right]\end{equation}
using the ansatz:
\[d{s^2} =  - {N^2}(t)d{t^2} + {a^2}(t)(dx_1^2 + dx_2^2 + dx_3^2)\]
\begin{equation}\phi  = \phi (t)\end{equation}
Then the Lagrangian becomes:
\begin{equation}L =  - v3a\frac{{{{\dot a}^2}}}{N} + v{a^3}\frac{1}{2}\frac{{{{\dot \phi }^2}}}{N} - v\Lambda N{a^3}\end{equation}
For the free minimally coupled scalar field one can introduce the change of variables:
\[T = V\cosh \frac{{\sqrt 3 }}{2}\phi \]
\begin{equation}X = V\sinh \frac{{\sqrt 3 }}{2}\phi \end{equation}
where $V = a^3$ and the Hamiltonian constraint becomes simply:
\begin{equation} - P_T^2 + P_X^2 + 4\Lambda = 0 \end{equation}
The solutions to the Wheeler-DeWitt equation in the $(V, \phi)$ variables are:
\begin{equation}\psi (V,\phi ) = {V^{-1}\chi _k}(V){e^{ik\phi }}\end{equation}
where:
\begin{equation}{{\bar \chi }_k} = {\left[ {(2/\pi )\sinh (\pi k)} \right]^{ - 1/2}}{J_{ - ik}}(\sqrt \Lambda  V)\end{equation}
or:
\begin{equation}{\chi _k} = \frac{1}{2}{(\pi )^{1/2}}{e^{\pi k/2}}H_{ik}^{(2)}(\sqrt \Lambda  V)\end{equation}
These two solutions are connected through a Bogoliubov transformation \cite{Birrell:1982ix}
\cite{Chitre:1977ip}:
\begin{equation}{{\bar \chi }_k} = {\alpha _k}{\chi _k} + {\beta _k}\chi _k^ * \end{equation}
where:
\[{\alpha _k} = {\left[ {{e^{\pi k}}/2\sinh (\pi k)} \right]^{1/2}}\]
\begin{equation}{\beta _k} = {\left[ {{e^{ - \pi k}}/2\sinh (\pi k)} \right]^{1/2}}\end{equation}
The propagator or Greens function in the $(V,\phi)$ variables is then of the form:
\begin{equation}
G(V,\phi ,V',\phi ') = \frac{1}{2\pi }\int dk {e^{i k(\phi  - \phi ')}\frac{i\pi }{2V V'} H_{i k}^{(2)}(\sqrt \Lambda  V_ > )J_{i k}(\sqrt \Lambda  {V'}_<)}
\end{equation}
This can be written as a integral over a Bessel function as:
\begin{equation}G(V,\phi ,V',\phi ') = \frac{1}{{2\pi }}\int {dk{e^{ik(\phi  - \phi ')}}\int_0^\infty  {\frac{{d\tau }}{\tau }} {e^{ - i\Lambda \tau }}{e^{({V^2} + V{'^2})/4i\tau }}} I_{ik}^{(2)}\left( {\frac{{iVV'}}{{2\tau }}} \right)\end{equation}
Defining the Kernel function $K(V,\phi,V';V,\phi')$ as:
\begin{equation}G(V,\phi ,V',\phi ') = \int_0^\infty  {\frac{{d\tau }}{\tau }} {e^{ - i\Lambda \tau }}K(V,\phi ,V',\phi ';\tau )\end{equation}
we have:
\begin{equation}K(V,\phi ,V',\phi ';\tau ) = \frac{1}{{2\pi }}\int {dk{e^{ik(\phi  - \phi ')}}{e^{({V^2} + V{'^2})/4i\tau }}I_{ik}^{(2)}\left( {\frac{{iVV'}}{{2\tau }}} \right)} \end{equation}
In terms of the $(X,T)$ variables the solutions to the Wheeler-DeWitt equation are:
\begin{equation}\psi (T,X) = \frac{1}{{\sqrt {2{{\left( {{K^2} + \Lambda } \right)}^{1/2}}} }}{e^{ - iT{{\left( {{K^2} + \Lambda } \right)}^{1/2}} + i K X}}\end{equation}
The propagator in these variables is:
\begin{equation}G(T,X,0,0) = \frac{1}{{2\pi }}\int {dK} \frac{1}{{2{{\left( {{K^2} + \Lambda } \right)}^{1/2}}}}{e^{ - i\left| T \right|{{\left( {{K^2} + \Lambda } \right)}^{1/2}} + i K X}}\end{equation}
which after performing the integrals gives \cite{Zhang:2008jy}:
\begin{equation}G(T,X,0,0) = 
\theta \left( {{T^2} - {X^2}} \right)\left( { - \frac{i}{4}H_0^{(2)}\left( {\sqrt \Lambda  \sqrt {{T^2} - {X^2}} } \right)} \right) + \theta \left( {{X^2} - {T^2}} \right)\left( {\frac{1}{{2\pi }}{K_0}\left( {\sqrt \Lambda  \sqrt {{X^2} - {T^2}} } \right)} \right)\end{equation}
This can also be written as:
\begin{equation}G(T,X,0,0) = \int_0^\infty  {\frac{{d\tau }}{\tau }} {e^{ - i\Lambda \tau }}{e^{ - \frac{{{X^2} - {T^2}}}{{2i\tau }}}}\end{equation}
so the Kernel function is simple in this case. The Green's function is related to the Kernel through the integral over a proper or fifth time \cite{Brown:1990iv}
\cite{Teitelboim:1980my}
\cite{Henneaux:1981su}
\cite{Ambjorn:1990jh}
\cite{Greensite:1991qt}
\cite{Garay:1990re}
\cite{Halliwell:1986ja}. Because of its simplicity it is possible to map the Kernel function in this case to the product of two Kernel functions of a free particle moving on a line. This Kernel can be computed using the EOH quantum algorithm in QISkit applied to the computation of the path integral between two points on a quantum computer with no potential. The setup is similar to the VQE in that one inputs a discretized Hamiltonian in terms of Pauli terms into the EOH algorithm in QISKit. The difference is  one does not need a variational ansatz or optimizer for the computation. The algorithm uses the Trotter-Suzuki approximation where one breaks up the Hamiltonian evolution into a product over time steps and expands each of these exponentials  in a series for small time steps. Finally one represents this expansion in terms of quantum gates in a quantum circuit through the Pauli term expansion. The results of the computation is shown in figure 13. For this computation we used the finite difference basis and 5 qubits. The multiple curves in time indicate $K(X,0;\tau)$ labelled by $X$. This computation agrees closely with the exact result for the path integral on a line with no potential. One can improve the agreement by increasing the number of time slices in the path integral computation in the Trotter-Suzuki approximation. This has the effect of generating a larger depth quantum circuit for the computation and increasing the running time on the quantum simulator.
\begin{figure}
    \centering
      \includegraphics[width=.7\linewidth]{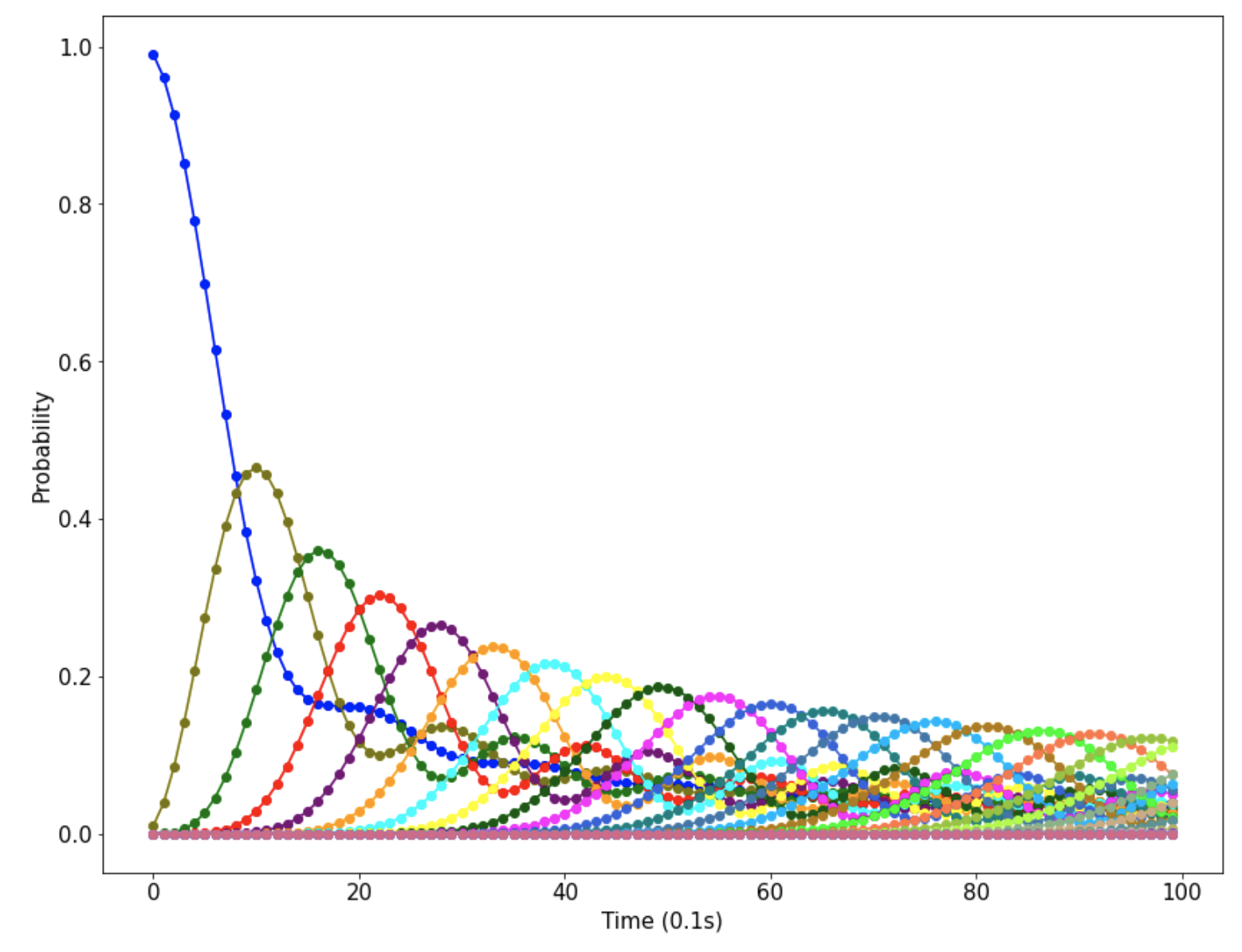}
    \caption{Quantum computation of magnitude squared of the Kernel for propagation along an interval with 32 vertices.}
    \label{3-qubit_unit}
\end{figure}

\subsection*{Three dimensional gravity with $T^2$ spatial topology}
Three dimensional gravity is a good model to study as it does not contain gravitons and thus is simpler than four dimensional gravity. For three dimensional gravity one defines:
\begin{equation}d{s^2} =  - {N^2}d{t^2} + {h_{ij}}d{x^i}d{x^j}\end{equation}
parametrizing the spatial two-metric as in \cite{Waldron:2004gg} :
\begin{equation}({h_{ij}}) = \left( {\begin{array}{*{20}{c}}
{T + X}&Y\\
Y&{T - X}
\end{array}} \right)\end{equation}
and these variables the Hamiltonian constraint is:
\begin{equation} - P_T^2 + P_X^2 + P_Y^2 + 4\Lambda  = 0\end{equation}
In terms of these variables the solution to the Wheeler-DeWitt equation is:
\begin{equation}\psi (T,X) = \frac{1}{{\sqrt {2{{\left( {P_x^2 + P_y^2 + 4\Lambda } \right)}^{1/2}}} }}{e^{ - iT{{\left( {P_x^2 + P_y^2 + 4\Lambda } \right)}^{1/2}} + i{P_x}X + i{P_y}Y}}\end{equation}
The Greens function is given by \cite{Zhang:2008jy}:
\[G(T,X,Y,0,0,0) = \frac{1}{{{{\left( {2\pi } \right)}^2}}}\int {d{P_x}d{P_y}} \frac{1}{{2{{\left( {P_x^2 + P_y^2 + 4\Lambda } \right)}^{1/2}}}}{e^{ - i\left| T \right|{{\left( {P_x^2 + P_y^2 + 4\Lambda } \right)}^{1/2}} + i{P_x}X + i{P_y}Y}}=\]
\begin{equation}\theta \left( {{T^2} - {X^2} - {Y^2}} \right)\left( { - i\frac{{{e^{ - i2\sqrt {\Lambda \left( {{T^2} - {X^2} - {Y^2}} \right)} }}}}{{4\pi \sqrt {\left( {{T^2} - {X^2} - {Y^2}} \right)} }}} \right) + \theta \left( {{X^2} + {Y^2} - {T^2}} \right)\left( {\frac{{{e^{ - 2\sqrt {\Lambda \left( {{X^2} + {Y^2} - {T^2}} \right)} }}}}{{4\pi \sqrt {\left( {{X^2} + {Y^2} - {T^2}} \right)} }}} \right)\end{equation}
The Kernel in this case is also simple:
\begin{equation}G(T,X,Y,0,0,0) = \int_0^\infty  {\frac{{d\tau }}{{{\tau ^{3/2}}}}} {e^{ - i4\Lambda \tau }}{e^{ - \frac{{{X^2} + {Y^2} - {T^2}}}{{2i\tau }}}}\end{equation}
The quantum computation of the Greens function is similar to the minimal scalar field case described above. One uses the EOH algorithm for the propagation of a free particle in fifth time $\tau$ and then integrates the product of three of these over $\tau$ to obtain the Greens function.

\subsection*{Three dimensional gravity with $S^2$ spatial topology}
Three dimensional gravity with $S^2$ topology is somewhat more complicated than $T^2$ topology because of the curvature.. In this model the Lagrangian is:
\begin{equation}S = \int {{d^3}x\sqrt { - g} } \left[ {\frac{1}{{16\pi G}}\left( {R - 2\Lambda } \right) - \frac{1}{2}{g^{\mu \nu }}{\partial _\mu }\phi {\partial _\nu }\phi } \right]\end{equation}
Using the metric ansatz:
\begin{equation}d{s^2} =  - N{(t)^2}d{t^2} + {a^2}(t)d\Omega _2^2\end{equation}
The Lagrangian becomes:
\begin{equation}L =  - v\frac{{{{\dot a}^2}}}{N} + vNk - vN{a^2}\Lambda  + v{a^2}\frac{{{{\dot \phi }^2}}}{{2N}}\end{equation}
In this case $v=4\pi$ the volume of a unit two sphere. Defining $a = e^\alpha$ this can be written as:
\begin{equation}L =  - v{a^2}\frac{{{{\dot \alpha }^2}}}{N} + v\frac{N}{{{a^2}}}{e^{2\alpha }}k - v\frac{N}{{{a^2}}}{e^{4\alpha }}\Lambda  + v{a^2}\frac{{{{\dot \phi }^2}}}{{2N}}\end{equation}
Now choosing the gauge $N=a^2$ this simplifies to:
\begin{equation}L =  - v{{\dot \alpha }^2} + v{e^{2\alpha }}k - v{e^{4\alpha }}\Lambda  + v\frac{{{{\dot \phi }^2}}}{2}\end{equation}
The Hamiltonian is then:
\begin{equation}H =  - \left( {v{{\dot \alpha }^2} + v{e^{2\alpha }}k - v{e^{4\alpha }}\Lambda  - v\frac{{{{\dot \phi }^2}}}{2}} \right)\end{equation} an in terms of canonical momentum this is:
\begin{equation}H =  - \frac{1}{{2v}}\left( {\frac{1}{2}p_\alpha ^2 + 2{v^2}{e^{2\alpha }}k - 2{v^2}{e^{4\alpha }}\Lambda  - p_\phi ^2} \right)\end{equation}
This is of the form of the Hamiltonian of a particle moving in an inverse Morse potential \cite{Halliwell:1989pu}
\cite{Duru:1984dx}. We plot the solution to the Wheeler-DeWitt equation and potential in figure 16.
\begin{figure}
    \centering
    \begin{minipage}[b]{0.7\linewidth}
      \centering
      \includegraphics[width=\linewidth]{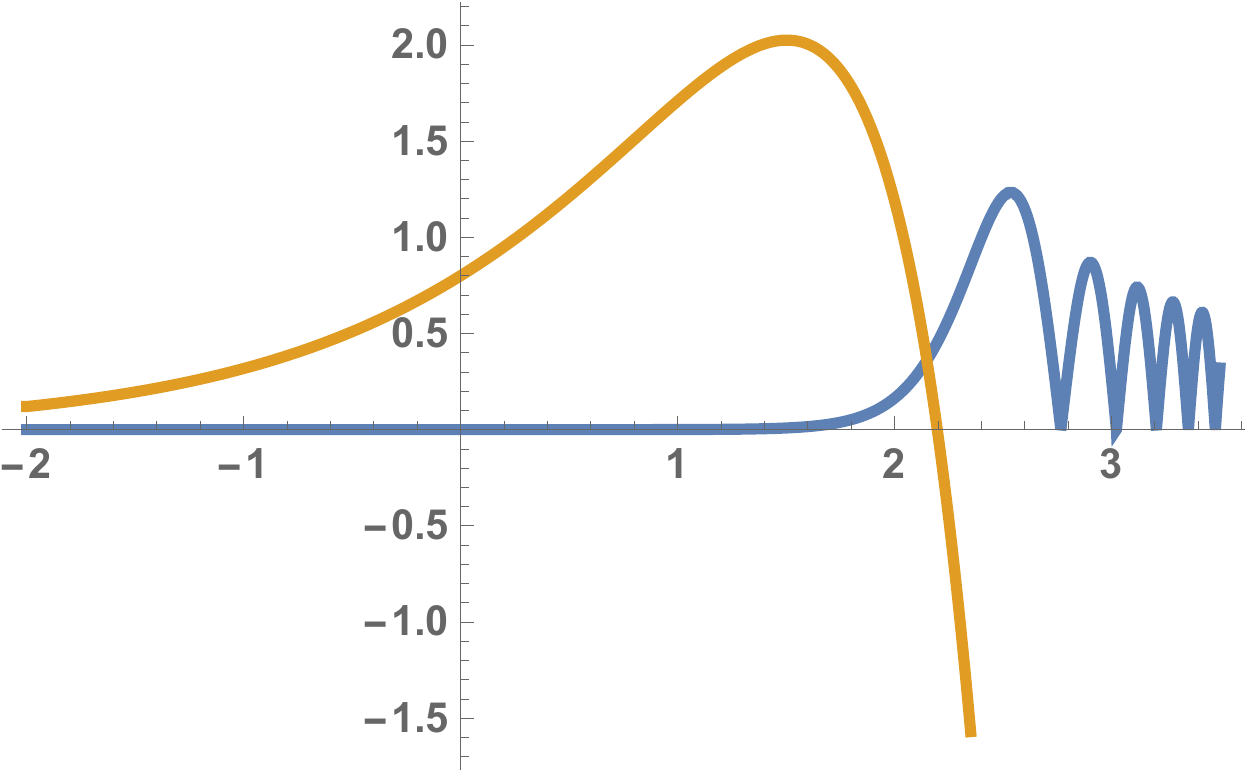}
    \end{minipage}
    \caption{Magnitude of wave function blue) and effective potential (orange) for three dimensional gravity with positive cosmological constant and spatial topology $S^2$.}
    \label{3-qubit_unit}
\end{figure}
Using the path integral for the Morse potential the Green's function in this case can be written as:
\begin{equation}G(\alpha ,\phi ,\alpha ',\phi ') = \frac{1}{{2\pi }}\int {dk{e^{ik(\phi  - \phi ')}}\int_0^\infty  {\frac{{d\tau }}{{\sinh (\sqrt \Lambda  \tau )}}} {e^{ - i\Lambda \tau }}{e^{i\frac{{\sqrt \Lambda  }}{2}({e^{4\alpha }} + {e^{4\alpha '}})\coth (\sqrt \Lambda  \tau )}}} {I_{ik}}\left( {\frac{{i{e^{2\alpha }}{e^{2\alpha '}}}}{{\sinh (\sqrt \Lambda  \tau )}}} \right)\end{equation}
which takes a similar form to the minimally coupled scalar field discussed above.
 For positive $\Lambda$ the Hamiltonian is not bounded below and one cannot use the VQE algorithm. It is possible to use the EOH algorithm however and one can consider Hamiltonian evolution a fifth time $\tau$ and one integrates over $\tau$ as a final step. For three dimensional gravity with positive cosmological constant the potential is the same form as the inverted Morse potential as in figure 14. In figure 15 we show the result of the EOH computation for three values of the fifth time parameter $\tau$. The calculation is similar to a tunnelling computation and is related to the tunnelling form of the wave function \cite{Kramer} used by \cite{Vilenkin:1987kf} or for the wave function used in  string cosmology \cite{Gasperini:2021eri}.
\begin{figure}[htb!]
\centering
\includegraphics[scale=0.7]{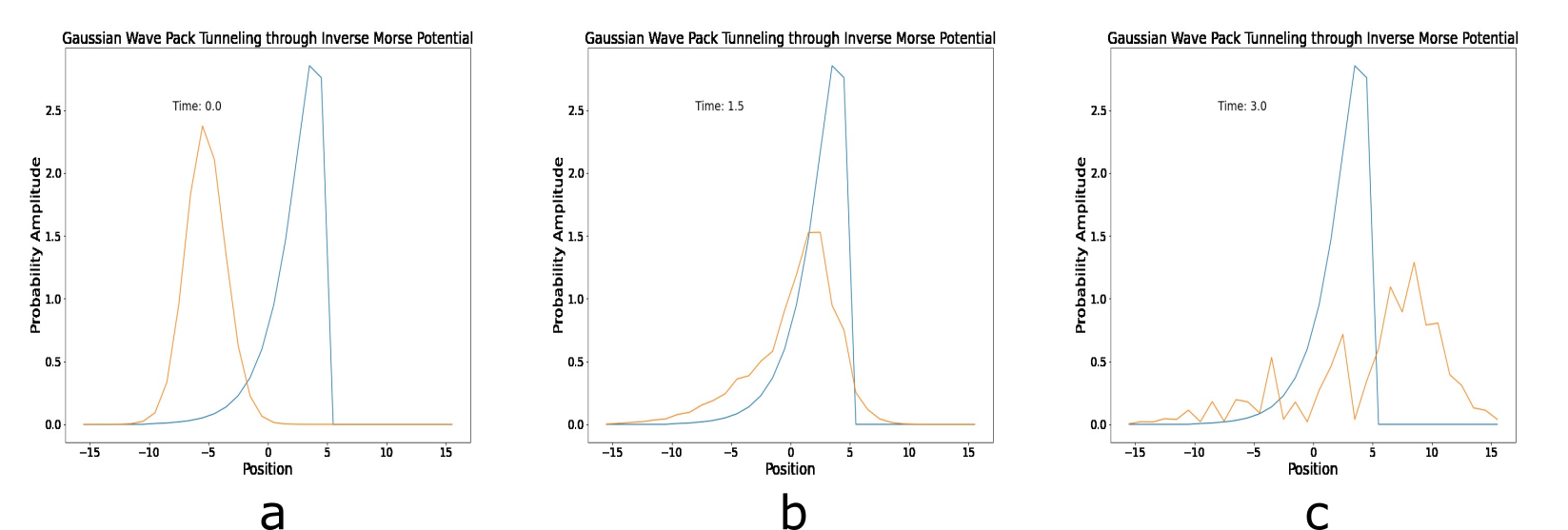}
\caption{Kernel computation done in Qiskit using the Trotter-Suzuki method. a) Kernel function for initial Gaussian state at small value of fifth time  b) Kernel function at intermediate fifth time c) Kernel function at large fifth time.}
\label{fig:tunnelingInverseMorse}
\end{figure}

%\subsection{Mapping Quantum cosmology to free particle path integral}

In some cases one can map the mini-superspace cosmology to the free particle path integral. This occurs when one can make a change of variables to the mini-superspaces metric to the Minskowki metric. some examples are gravity coupled to a scalar field and three dimensional gravity with $T^2$ spatial topology. These cases are the most straightforward to simulate on a quantum computer because the mini-superspace metric is flat. Other case can be treated on the quantum computer by mapping the system onto a particle moving in a potential.

\subsection*{Two dimensional quantum gravity with $S^1$ spatial topology}

Wheeler-DeWitt approaches can be applied to two dimensional gravity models such as JT gravity or Liouville gravity \cite{Casali:2021ewu}
\cite{Betzios:2020nry}
\cite{Iliesiu:2020zld}
\cite{Martinec:1984fs}
\cite{Witten:2020wvy}
\cite{Maldacena:2005he}
\cite{Douglas:2003up} however unlike the models described above one has a complete description of the path integral as a Matrix model. As an example for the Super-Liouville theory couple to a scalar field we have \cite{Douglas:2003up}:
\[\left( { - \frac{{{\partial ^2}}}{{\partial {\phi ^2}}} \pm {b^2}\mu {e^{b\phi }} + {b^2}{\mu ^2}{e^{2b\phi }} - {b^2}{p^2}} \right){\psi _{p \pm }}(\phi ) = 0\]
\begin{equation}\left( { - \frac{{{\partial ^2}}}{{\partial {\phi ^2}}} + {b^2}{\mu ^2}{e^{2b\phi }} - {b^2}{p^2}} \right){\psi _{p0}}(\phi ) = 0\end{equation}
for the RR sector where left and right movers are periodic. 
In the NS-NS sector we have:
\[{\psi _{p + }}(\phi ) = \frac{p}{{\sqrt 2 }}{e^{ - b\phi /2}}{W_{ - \frac{1}{2},ip}}(2{e^{b\phi }}) =    - i\frac{1}{{\sqrt {2\pi } }}{e^{b\phi /2}}\left( {{K_{ip + \frac{1}{2}}}({e^{b\phi }}) - {K_{ip - \frac{1}{2}}}({e^{b\phi }})} \right)\]
\[{\psi _{p - }}(\phi ) = \frac{1}{{\sqrt 2 }}{e^{ - b\phi /2}}{W_{\frac{1}{2},ip}}(2{e^{b\phi }}) =   \frac{1}{{\sqrt {2\pi } }}{e^{b\phi /2}}\left( {{K_{ip + \frac{1}{2}}}({e^{b\phi }}) + {K_{ip - \frac{1}{2}}}({e^{b\phi }})} \right)\]
\begin{equation}{\psi _{p0}}(\phi ) = \frac{1}{{\sqrt 2 }}{e^{ - b\phi /2}}{W_{0,ip}}(2{e^{b\phi }}) =   \frac{1}{{\sqrt \pi  }}{K_{ip}}({e^{b\phi }})\end{equation}
These two cases are related to the Morse potential and exponential potential. For two dimensional quantum gravity such as the Liouville model or JT gravity one can represent the partition function of the theory in terms of Matrix integrals so one can have precise checks of the quantum computer simulations for these cases.

\subsection*{Negative cosmological constant}

Another fortuitous case is a negative cosmological constant discussed in \cite{Kocher:2018ilr}. Here because the gravitational potentials are right side up the VQE quantum algorithm can be used to calculate the solutions to the Wheeler-DeWitt equation. For example for three dimensional gravity with negative spatial curvature negative and negative cosmological constant the Wheeler-DeWitt equation is:
\begin{equation}H \psi_p(\alpha) =  - \frac{1}{{2v}}\left( {-\frac{1}{2}\partial_\alpha ^2 - 2{v^2}{e^{2\alpha }}(-k) + 2{v^2}{e^{4\alpha }}(-\Lambda)  - p_\phi ^2} \right)\psi_p(\alpha)=0\end{equation}
This form is the same form as the Morse potential which can be  accurately simulated on a quantum computer. One can also use the EOH algoritm to compute the Green's function for negative cosmological constant. In this case the potential can be written as double well potential. The result of the EOH computation at various fifth times are shown in figure 16. The Green's function is related to the integral of the Kernel function over the fifth time.
Finally for the case of negative $\Lambda$ and Anti-deSitter Space (AdS) these solution can be related to  calculations in a Conformal Field Theory (CFT) through the AdS/CFT correspondence.
\begin{figure}
    \centering
    \includegraphics[scale=0.4]{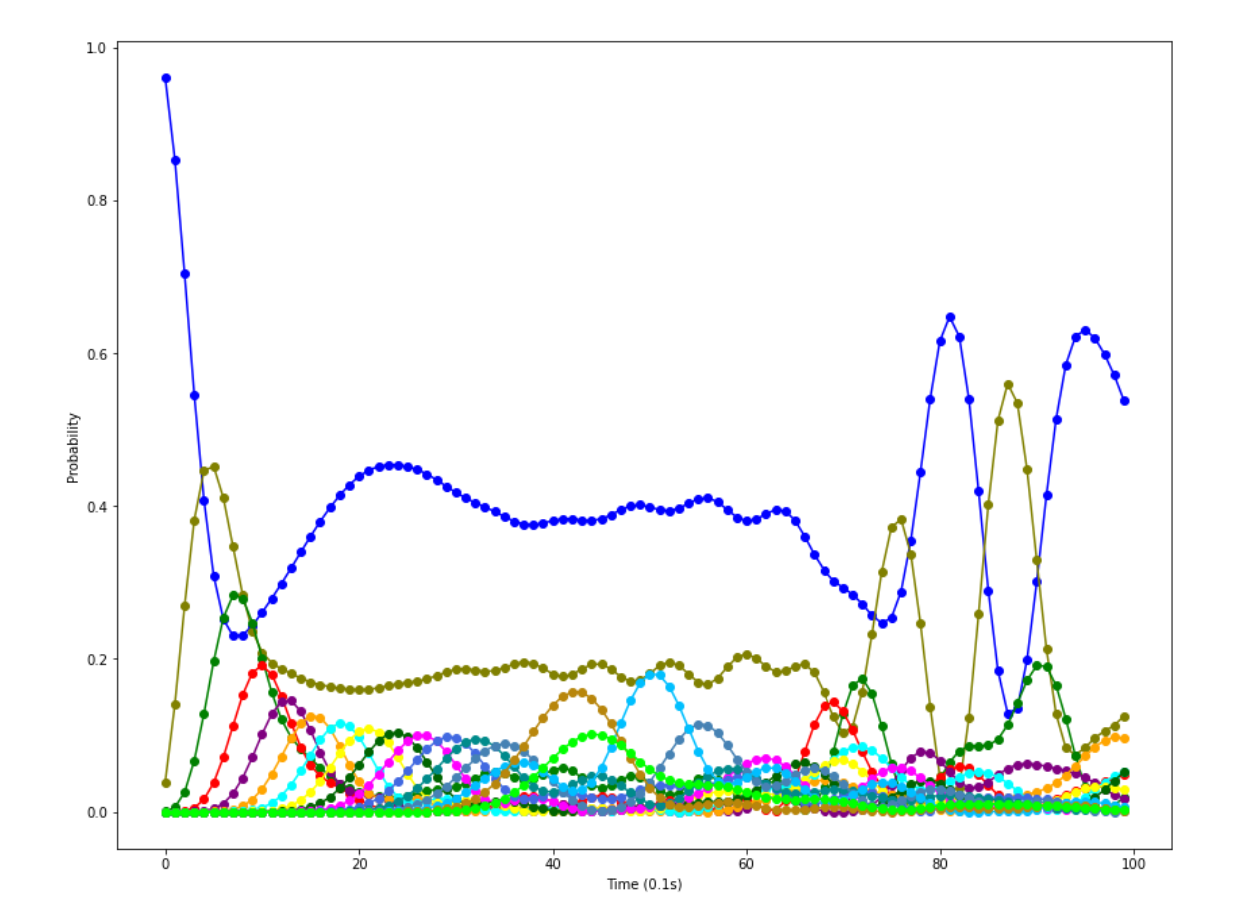}
    \caption{Result of EOH quantum computation for double well potential associated to quantum cosmology with negative cosmological constant.}
    \label{fig:mesh1}
\end{figure}
\subsection*{Inhomogeneous models}

Besides the homogeneous models studied above one can also consider inhomogeneous models such as those considered in 
\cite{Misner:1973zz}
\cite{FernandoBarbero:2010qy}
\cite{McGuigan:1990nd}
\cite{Fischler:1989se}
\cite{Berger:1972pg}
\cite{Berger:1975kn}. These models are much more complex than the homogeneous or minisuperspace examples discussed above and are referred to as midi-suoerspace models. Examples of these type of models are the Spherical-$S^2$ model and Gowdy-$T^2$ model where one compactifies four dimensional gravity on $S^2$ or $T^2$ respectively. 
Here the ansatz depends on two variables $(t,r)$ and is expressed as:
\begin{equation}d{s^2} =  - {N^2}d{t^2} + {a^2}(t,r){\left( {dr + {N_1}dt} \right)^2} + {b^2}(t,r)d\Omega _2^2\end{equation}
A form which is somewhat similar to the Kaluza Klein Cosmology above. Varying with respect to $N$ yields the Hamiltonian constraint and varying with respect to $N_1$ yields the momentum constraint.
In terms of the variables $a=e^\alpha$ and $b=e^\beta$ these become:
\begin{equation}H = \frac{{p_\alpha ^2}}{4} - \frac{{{p_\alpha }{p_\beta }}}{2} + \frac{{p_\phi ^2}}{2} + c_2^2{e^{4\beta }}\left( {3{{\left( {\beta '} \right)}^2} + 2\beta '' + 2\alpha '\beta ' + {{\left( {\phi '} \right)}^2}} \right) + c_2^2{e^{2\alpha }}{e^{4\beta }}( -e^{-2\beta}  + \lambda + V(\phi ))\end{equation}
and
\begin{equation}P = \beta '{p_\beta } - \alpha '{p_\alpha } + \phi '{p_\phi }\end{equation}
In this form one can see the resemblance to the string theory constraints involving the $L_0+\bar L_0$ and  $L_0-\bar L_0$ operators. Indeed one expects the quantization of these models to be at least as complex as Light-cone String field quantization in a time dependent background. Despite or perhaps because of this complexity these models represent an opportunity for a quantum speedup over classical computers because of their unique capability to perform Lorentzian path integrals without relating them to Euclidean path integrals.

 Finally more representations of Hamiltonian constraints are contained in the Appendix  in terms of various choices of variables and gauge fixing condition involving the lapse function and the spatial metric.

\section{ Conclusion}

In this paper we applied quantum computing methods such as the VQE and EOH quantum algorithms to inflationary, dark energy, dark matter cosmological models and quantum cosmology. The results of the quantum computations agreed quite well with the classical computations at least for the state vector simulator that we used in this paper. It will be important to see how noise affects the results either using the QASM simulator or Noisy Intermediate Scale Quantum Computers (NISQ). Potential areas of quantum advantage are in calculating Green's functions for inhomogeneous models and computing path integrals for higher dimensional models, where one has a large number of interacting boson fields. Even for the simple case of two dimensional inhomogeneous  models coupled to matter one has exponential number of quantum states which can  be efficiently simulated on a quantum computer and has memory constraints on classical computers. Our investigations indicate that the intersection of quantum computing and cosmology is an interesting new application area for quantum computing that can be explored in the near term era of quantum computation.

\section*{Acknowledgements}

 This material is based upon work supported in part by the U.S. Department of Energy, Office of Science, National Quantum Information Science Research Centers, Co-design Center for Quantum Advantage (C2QA) under contract number DE-SC0012704. This project was supported in part by the U.S. Department of Energy, Office of Science, Office of Workforce Development for Teachers and Scientists (WDTS) under the Science Undergraduate Laboratory Internships Program (SULI).

\section*{Appendix}

\subsection*{ Inverted Liouville potential}
defining  $ a = {e^{\alpha  }} $ this can be written as:
\begin{equation}L =  - v3{a^3}\frac{{{{\dot \alpha }^2}}}{N} + v{a^3}\frac{1}{2}\frac{{{{\dot \phi }^2}}}{N} - \Lambda \frac{N}{{{a^3}}}{e^{6\alpha }}\end{equation}
%\[L =  - v\frac{1}{2}{e^{3\alpha  }}\frac{{{{\dot \alpha }^2}}}{N} + v{e^{3\alpha /\sqrt 6 }}\frac{1}{2}\frac{{{{\dot \phi }^2}}}{N} - v\Lambda N{e^{3\alpha /\sqrt 6 }}\]
In this case $v=(2\pi)^3$ the volume of a unit three torus. Fixing the gauge $ N = a^3 $ this simplifies to:
\begin{equation}L =  - v3{{\dot \alpha }^2} + v{a^3}\frac{1}{2}{{\dot \phi }^2} - v\Lambda {e^{6\alpha }}\end{equation}
%\[L =  - \frac{1}{2}{{\dot \alpha }^2} + \frac{1}{2}{{\dot \phi }^2} - {v^2}\Lambda {e^{6\alpha /\sqrt 6 }}\]
With the Hamiltonian given by:
\begin{equation}H =  - \left( {\frac{1}{{12v}}{p_\alpha }^2 - \frac{1}{{2v}}p_\phi ^2 - v\Lambda {e^{6\alpha }}} \right)\end{equation}
%\[H =  - \frac{1}{2}{p_\alpha }^2 + \frac{1}{2}{p_\phi }^2 + {v^2}\Lambda {e^{6\alpha /\sqrt 6 }} = 0\]
or
\begin{equation}H =  - \frac{1}{{6v}}\left( {\frac{1}{2}{p_\alpha }^2 - 3p_\phi ^2 - {v^2}6\Lambda {e^{6\alpha }}} \right)\end{equation}
with canonical momentum $p_\alpha$ and $p_\phi$. The Wheeler DeWitt equation is then written as:
\begin{equation}H\psi (\alpha ,\phi ) = 0 \end{equation} or:
\begin{equation}\left( { - \partial _\alpha^2 + 6\partial _\phi^2 - 12{v^2}\Lambda {e^{6\alpha }}} \right)\psi (\alpha ,\phi ) = 0\end{equation}
%\[H\psi (\alpha ,\phi ) = \left( {\frac{1}{2}{\partial _\alpha }^2 - \frac{1}{2}{\partial _\phi }^2 + {v^2}\Lambda {e^{6\alpha /\sqrt 6 }}} \right)\psi (\alpha ,\phi ) = 0\]
The solutions are separable as  $\psi (\alpha ,\phi ) = {e^{i{p_\phi }\phi }}{\psi _{{p_\phi }}}(\alpha )$ and can be expressed in terms of Bessel functions. The form of the potential in the Hamiltonian is that of the inverse Liouville potential. We plot this potential and the wave function in figure 17. Note how the oscillation in wave function increases in positive $\alpha$ as the canonical momentum $p_\alpha$ is increasing there which is an aspect of the Universe accelerating due to positive $\Lambda$.

An important quantity to compute in quantum cosmology is the Green function $G(\alpha, \alpha', \phi, \phi')$ which is defined from the gravitational Lorentzian path integral. This is related to the integral over $\tau$ of a Kernel function given by:
\begin{equation}K(\alpha ,\alpha ',\phi ,\phi ';\tau) = \int {d{p_\phi }{e^{i{p_\phi }\phi }}{K_{{p_\phi }}}(} \alpha ,\alpha ';\tau)\end{equation}
The Kernel function can be determined from the Hamiltonian through.
\begin{equation}{K_{{p_\phi }}}(\alpha ,\alpha ';\tau) = \left\langle \alpha  \right|{e^{ - iH \tau}}\left| {\alpha '} \right\rangle  = \left\langle \alpha  \right|U(\tau)\left| {\alpha '} \right\rangle \end{equation}
where
\begin{equation}U(\tau) = {e^{ - iH \tau}}\end{equation}
$U(\tau)$ is a unitary operator which can be represented on a quantum computer using quantum gates and quantum circuits. 

\begin{figure}
    \centering
    \begin{minipage}[b]{0.7\linewidth}
      \centering
      \includegraphics[width=\linewidth]{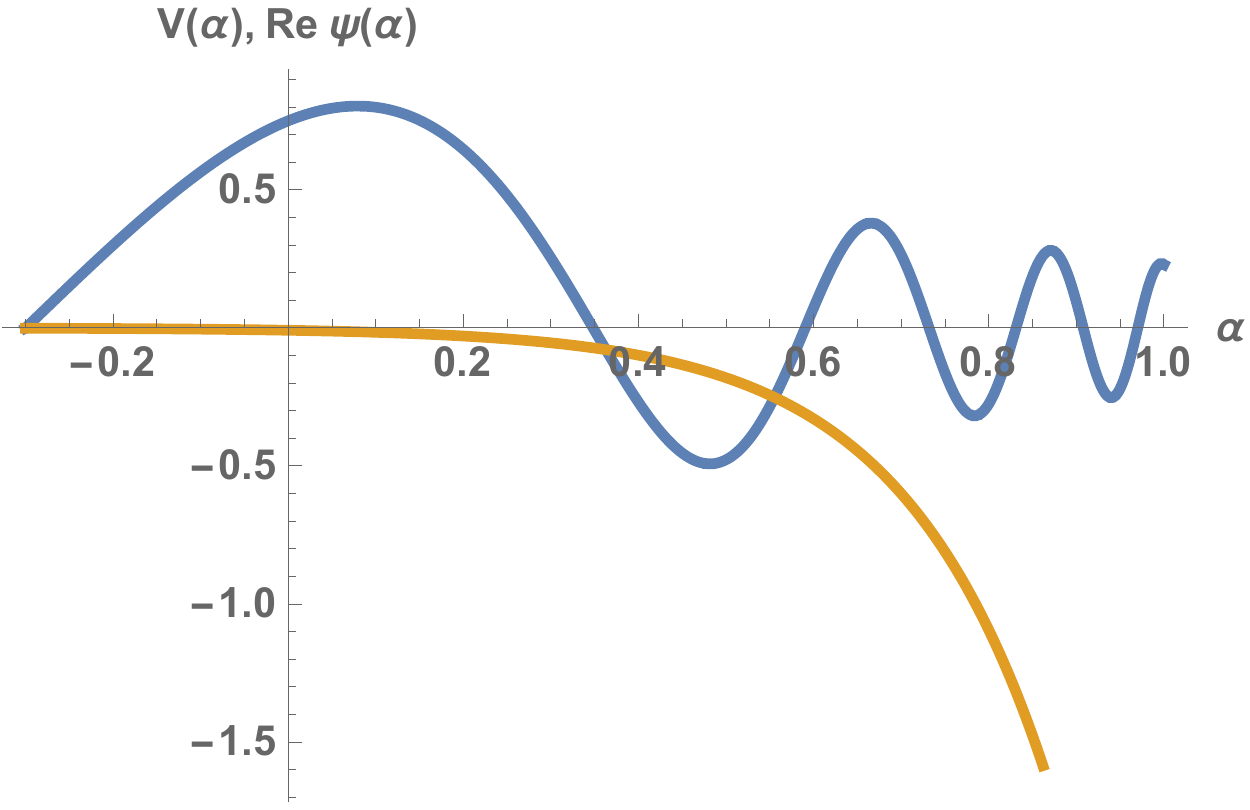}
    \end{minipage}
    \caption{Real part of ave function (blue) and effective inverse exponential potential (orange) for  gravity with positive cosmological constant and spatial topology $T^3$ of radius  $a = e^{\alpha}$.}
    \label{3-qubit_unit}
\end{figure}

\subsection*{Inverted oscillator potential}
If one defines  $y= a^{3/2}$ the mini-superspace Lagrangian  can be written as:
\begin{equation}L =  - v\frac{4}{3}\frac{{{{\dot y}^2}}}{N} + v{y^2}\frac{1}{2}\frac{{{{\dot \phi }^2}}}{N} - v\Lambda N{y^2}\end{equation}
The canonical momentum are then given by:
\[{p_y} =  - v\frac{8}{3}\frac{{\dot y}}{N}\]
\begin{equation}{p_\phi } = v\frac{{{{\dot \phi }^2}}}{N}\end{equation}
The Hamiltonian constraint is derived from varying $L$ with respect to $N$ and is:
\begin{equation}H = -\frac{3}{{16v}}p_y^2 + \frac{1}{{2v}}\frac{{p_\phi ^2}}{{{y^2}}} + v\Lambda {y^2}\end{equation}
Choosing the gauge $N=1$ the Wheeler-DeWitt equation is given by:
\begin{equation}H\psi  = -\left( {\frac{3}{{16v}}p_y^2 - \frac{1}{{2v}}\frac{{p_\phi ^2}}{{{y^2}}} - v\Lambda {y^2}} \right)\psi  = 0\end{equation}
or:
\begin{equation}H\psi  = -\frac{3}{{8v}}\left( {\frac{1}{2}p_y^2 - \frac{4}{3}\frac{{p_\phi ^2}}{{{y^2}}} - {v^2}\frac{8}{3}\Lambda {y^2}} \right)\psi  = 0\end{equation}
\begin{figure}
    \centering
    \begin{minipage}[b]{0.7\linewidth}
      \centering
      \includegraphics[width=\linewidth]{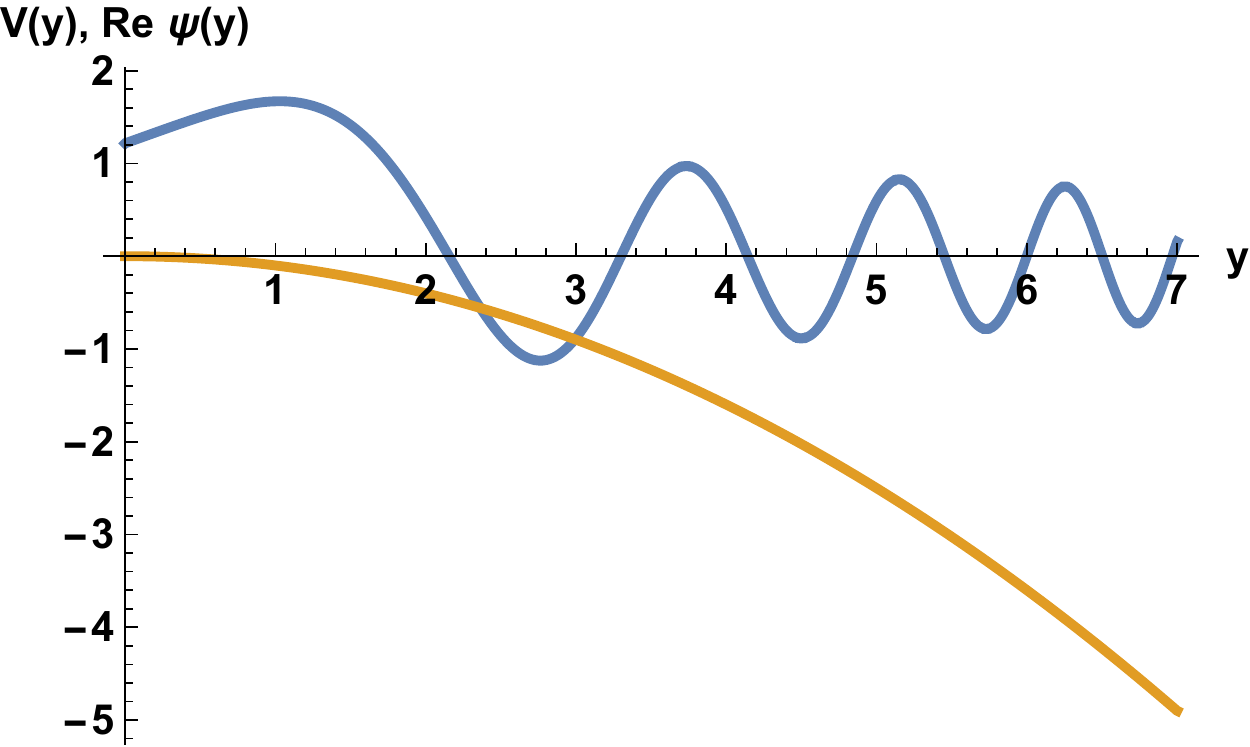}
    \end{minipage}
    \caption{Real part of ave function (blue) and effective inverse quadratic potential (orange) for  gravity with positive cosmological constant and spatial topology $T^3$ of radius to the three halves   $y = a^{3/2}$.}
    \label{3-qubit_unit}
\end{figure}
The formula of the Kernel for the inverted harmonic oscillator is:
\begin{equation}K(y,|y',\tau) = \sqrt {\frac{{\omega }}{{2\pi i\sinh (\omega \tau)}}} \exp \left[ {\left( {\frac{{i \omega }}{{2\sinh (\omega \tau)}}} \right)\left( {\left( {{y^2} + y{'^2}} \right)\cosh (\omega \tau) - 2yy'} \right)} \right]\end{equation}
The Greens function is then given by:
\begin{equation}G(y,y') = \int_0^\infty  {K(y} ,\tau|y',0)d\tau\end{equation}
This can be written as:
%\[\psi (y,t) = \int {dy'} K(y,t|y',t')\psi (y',t')\]
\begin{equation}G(y,y') = {\psi _i}({y_ < })\psi _f^ * ({y_ > })\end{equation}
where $\psi_i(y)$ and $\psi_f(y)$ satisfy the Wheeler-DeWitt equation. The solution and effective potential are plotted in figure 18.

\subsection*{Inverted linear potential}

Choosing variables:
\begin{equation}x = {a^2}\end{equation}
The Einstein-Hilbert Lagrangian coupled to a scalar field becomes:
\begin{equation}L =  - v\frac{3}{4}\frac{{{{\dot x}^2}}}{{Na}} + v3Nak + v{x^2}\frac{1}{2}\frac{{{{\dot \phi }^2}}}{{Na}} - v\Lambda Nax\end{equation}
Choosing the gauge $Na=1$ this simplifies to:
\begin{equation}L =  - v\frac{3}{4}{{\dot x}^2} + v3k + v{x^2}\frac{1}{2}{{\dot \phi }^2} - v\Lambda x\end{equation}
and the Hamiltonian is:
\begin{equation} H = -\frac{1}{{3v}}p_x^2 - v3k + \frac{1}{{2v{x^2}}}p_\phi ^2 + v\Lambda x\end{equation}
which can be written as:
\begin{equation}  H = -\frac{2}{{3v}}\left( {\frac{1}{2}p_x^2 + \frac{9}{2}{v^2}k - \frac{3}{4}\frac{{p_\phi ^2}}{{{x^2}}} - \frac{3}{2}{v^2}\Lambda x} \right)\end{equation}
and is of the form of a particle moving in a inverse linear potential. The Wheeler-DeWitt equation $H\psi =0$ can be solve for the case $p_\phi=0$ in terms of the Airy function \cite{Feldbrugge:2017kzv}
\cite{Caputa:2018asc}. We plot the potential and solution in figure 19.
\begin{figure}
    \centering
    \begin{minipage}[b]{0.7\linewidth}
      \centering
      \includegraphics[width=\linewidth]{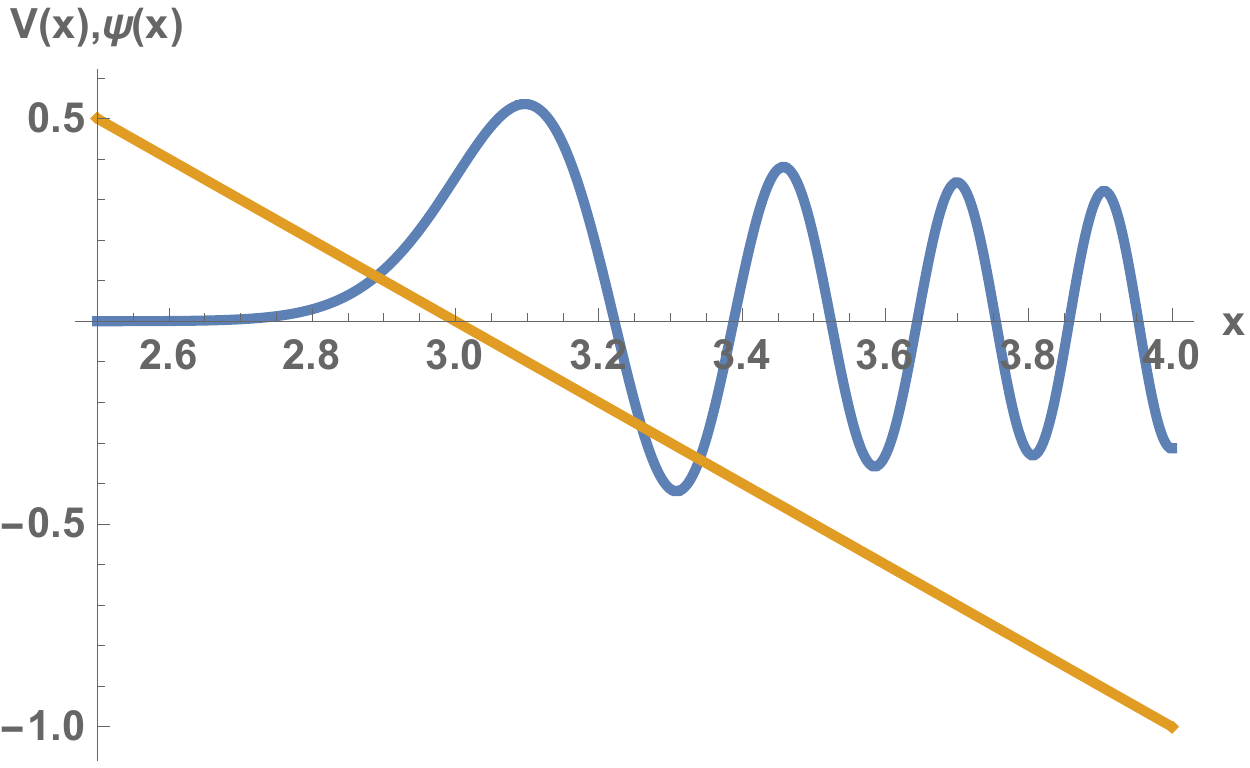}
    \end{minipage}
    \caption{Wave function (blue) and effective inverse linear potential (orange) for  gravity with positive cosmological constant and spatial topology $S^3$ of radius squared  $x = a^2$.}
    \label{3-qubit_unit}
\end{figure}
The Kernel function for the inverted linear potential is
\begin{equation}K(x,\tau|x',0) = \sqrt {\frac{1}{{2\pi i \tau}}} \exp \left[ {i\left( {\frac{1}{{2\tau}}{{\left( {x - x'} \right)}^2} + \frac{{f \tau }}{2}\left( {x + x'} \right) - \frac{{{f^2}{\tau^3}}}{{24}}} \right)} \right]\end{equation}
with $f = \frac{3}{2}{v^2}\Lambda$
\subsection*{Inverse quartic potential}

In the conformal gauge $N=a$ the gravity-matter Lagrangian becomes:
\begin{equation}L =  - v3{{\dot a}^2} + v3{a^2}k + v{a^2}\frac{1}{2}{{\dot \phi }^2} - v\Lambda {a^4}\end{equation}
In this case $v = 2\pi^2$ the volume of a unit three sphere. The Hamiltonian is:
\begin{equation}H =  - \left( {\frac{1}{{12v}}p_a^2 + 3vk{a^2} - \frac{{p_\phi ^2}}{{2v{a^2}}} - v\Lambda {a^4}} \right)\end{equation}
which can be written as:
\begin{equation}H =  - \frac{1}{{6v}}\left( {\frac{1}{2}p_a^2 + 18{v^2}k{a^2} - 3\frac{{p_\phi ^2}}{{{a^2}}} - 6{v^2}\Lambda {a^4}} \right)\end{equation}
In this form the Hamiltonian is of the form of a particle moving in an inverse quartic potential. We plot the solution to the Wheeler DeWitt equation and potential in figure 20.

\begin{figure}
    \centering
    \begin{minipage}[b]{0.7\linewidth}
      \centering
      \includegraphics[width=\linewidth]{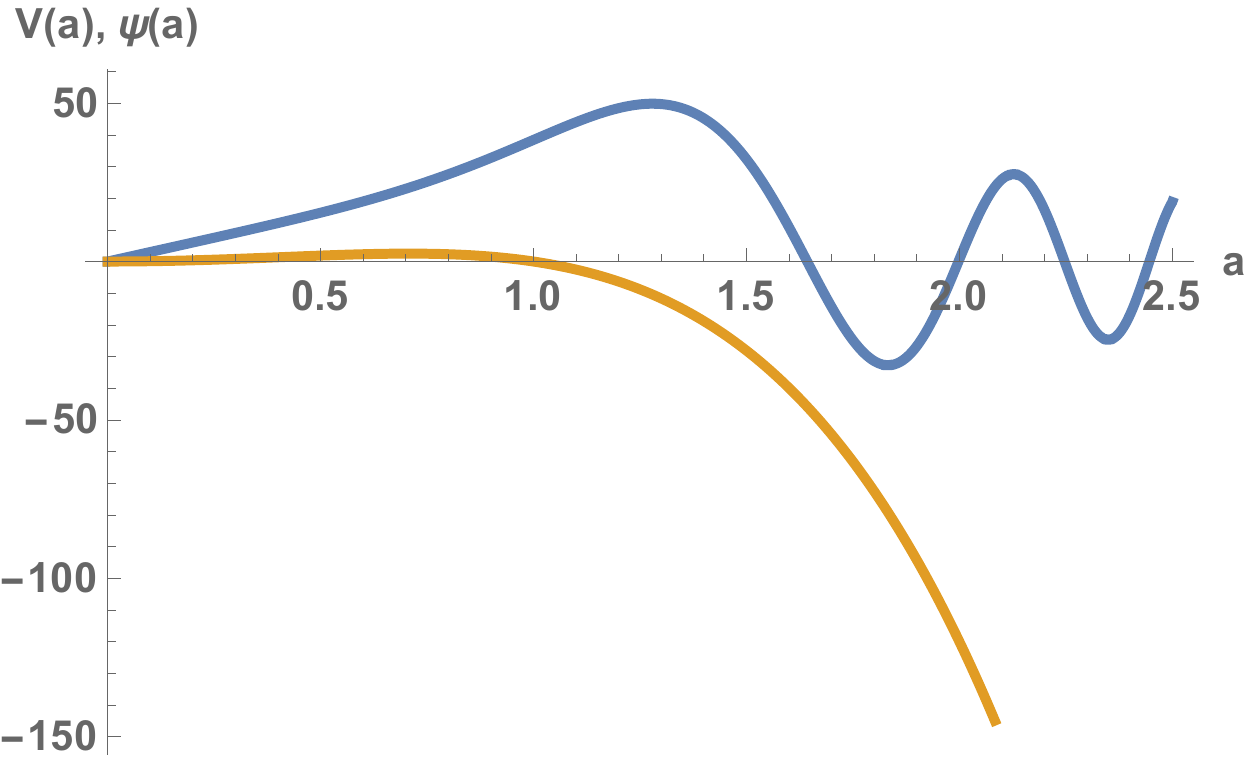}
    \end{minipage}
    \caption{Wave function (blue) and effective inverse quartic potential (orange) for  gravity with positive cosmological constant and spatial topology $S^3$ of radius $a$.}
    \label{3-qubit_unit}
\end{figure}

\subsubsection*{Kantowski-Sachs Cosmology}

For the Kanowski-Sachs cosmology the spatial sections have topology $S^1\times S^2$ and metric:
\begin{equation}d{s^2} =  - {N^2}d{t^2} + {a^2}d{x^2} + {b^2}d\Omega _2^2\end{equation}
The volume of $S^1 \times S^2$ is $c a b^2$ where $c= 8 \pi^2$. The Lagrangian then becomes:
\begin{equation}L = c\left[ { - \frac{{a{{\dot b}^2}}}{N} - 2\frac{{b\dot a\dot b}}{N} + Nka} \right]\end{equation}
Defining:
\[{e^q} = ab\]
\begin{equation}{e^\alpha } = a\end{equation}
we have:
\begin{equation}L = c\left[ { - {e^{2q}}\frac{{{{\dot q}^2}}}{{N{e^\alpha }}} + {e^{2q}}\frac{{{{\dot \alpha }^2}}}{{N{e^\alpha }}} + Nk{e^\alpha }} \right]\end{equation}
and the Hamiltonian constraint becomes:
\begin{equation}H =  - p_\alpha ^2 + p_q^2 + 4{c^2}k{e^{2q}} = 0\end{equation}
The Wheeler-DeWitt equation is:
\begin{equation}H\psi  = \left( {\partial _\alpha ^2 - \partial _q^2 + 4{c^2}k{e^{2q}}} \right)\psi  = 0\end{equation}
with solutions:
\begin{equation}\psi \left( {\alpha ,q} \right) = {K_{ip}}(2c{e^q})\end{equation}
where $K_\nu(z)$ is the modified Bessel function. Defining \cite{McGuigan:1993ma}:
\[X = {e^q}\cosh \alpha \]
\begin{equation}T = {e^q}\sinh \alpha \end{equation}
The Hamiltonian constraint is then:
\begin{equation}H =  - P_T^2 + P_X^2 + 4{c^2}k=0\end{equation}
and Wheeler-DeWitt equation becomes:
\begin{equation}H\psi  = \left( {\partial _T^2 - \partial _X^2 + 4{c^2}k} \right)\psi  = 0\end{equation}
with solutions:
\begin{equation}\psi (X,T) = \frac{1}{{\sqrt {2\left( {{P^2} + 4{c^2}k} \right)} }}{e^{ iXP - iT\sqrt {{P^2} + 4{c^2}k} }}\end{equation}
The Kernel and propagator for the Kantowski-Sachs cosmology  is thus similar to the scalar field case.

%\[b = {e^{ - \phi }}\]
%\[a = {e^\rho }\]
%We can form Minkowski coordinates in mini-superspace by:
%\[X = {e^q}\cosh \phi \]
%\[T = {e^q}\sinh \phi \]
%where $ q = \rho  - \phi $. The Hamiltonian constriant then becomes:
%\[ - P_T^2 + P_X^2 + \frac{1}{{2G}} = 0\]
%similar to the scalar field case above.


\begin{thebibliography}{0}


%\cite{Kocher:2018ilr}
%\cite{Ganguly:2019kkm}
%\cite{Li:2017gvt}
%\cite{Mielczarek:2018nnd}
%\cite{Mielczarek:2018jsh}
%\cite{Li:2020kbv}
%\cite{Antonini:2019qkt}
%\cite{Brown:2019hmk}
%\cite{Nezami:2021yaq}
\bibitem{Kocher:2018ilr}
C.~D.~Kocher and M.~McGuigan,
``Simulating 0+1 Dimensional Quantum Gravity on Quantum Computers: Mini-Superspace Quantum Cosmology and the World Line Approach in Quantum Field Theory,''
doi:10.1109/NYSDS.2018.8538963
[arXiv:1812.08107 [quant-ph]].

%\cite{Ganguly:2019kkm}
\bibitem{Ganguly:2019kkm}
A.~Ganguly, B.~K.~Behera and P.~K.~Panigrahi,
``Demonstration of Minisuperspace Quantum Cosmology Using Quantum Computational Algorithms on IBM Quantum Computer,''
[arXiv:1912.00298 [quant-ph]].

%\cite{Li:2017gvt}
\bibitem{Li:2017gvt}
K.~Li, Y.~Li, M.~Han, S.~Lu, J.~Zhou, D.~Ruan, G.~Long, Y.~Wan, D.~Lu and B.~Zeng, \textit{et al.}
``Quantum Spacetime on a Quantum Simulator,''
doi:10.1038/s42005-019-0218-5
[arXiv:1712.08711 [quant-ph]].

%\cite{Mielczarek:2018nnd}
\bibitem{Mielczarek:2018nnd}
J.~Mielczarek,
``Quantum Gravity on a Quantum Chip,''
[arXiv:1803.10592 [gr-qc]].

%\cite{Mielczarek:2018jsh}
\bibitem{Mielczarek:2018jsh}
J.~Mielczarek,
``Spin Foam Vertex Amplitudes on Quantum Computer - Preliminary Results,''
Universe \textbf{5}, no.8, 179 (2019)
doi:10.3390/universe5080179
[arXiv:1810.07100 [gr-qc]].

%\cite{Li:2020kbv}
\bibitem{Li:2020kbv}
Y.~Z.~Li and J.~Liu,
``On Quantum Simulation Of Cosmic Inflation,''
[arXiv:2009.10921 [quant-ph]].

%\cite{Antonini:2019qkt}
\bibitem{Antonini:2019qkt}
S.~Antonini and B.~Swingle,
``Cosmology at the end of the world,''
Nature Phys. \textbf{16}, no.8, 881-886 (2020)
doi:10.1038/s41567-020-0909-6
[arXiv:1907.06667 [hep-th]].

%\cite{Brown:2019hmk}
\bibitem{Brown:2019hmk}
A.~R.~Brown, H.~Gharibyan, S.~Leichenauer, H.~W.~Lin, S.~Nezami, G.~Salton, L.~Susskind, B.~Swingle and M.~Walter,
``Quantum Gravity in the Lab: Teleportation by Size and Traversable Wormholes,''
[arXiv:1911.06314 [quant-ph]].

%\cite{Nezami:2021yaq}
\bibitem{Nezami:2021yaq}
S.~Nezami, H.~W.~Lin, A.~R.~Brown, H.~Gharibyan, S.~Leichenauer, G.~Salton, L.~Susskind, B.~Swingle and M.~Walter,
``Quantum Gravity in the Lab: Teleportation by Size and Traversable Wormholes, Part II,''
[arXiv:2102.01064 [quant-ph]].

%\cite{Mocz:2021ehj}
\bibitem{Mocz:2021ehj}
P.~Mocz and A.~Szasz,
``Towards Cosmological Simulations of Dark Matter on Quantum Computers,''
Astrophys. J. \textbf{910}, no.1, 29 (2021)
doi:10.3847/1538-4357/abe6ac
[arXiv:2101.05821 [astro-ph.CO]].

\bibitem{Berger:1993fm}
B.~K.~Berger,
``Application of Monte Carlo simulation methods to quantum cosmology,''
Phys. Rev. D \textbf{48}, 513-529 (1993)
doi:10.1103/PhysRevD.48.513


%\cite{Kandala}
\bibitem{Kandala} Kandala, Abhinav, et al. "Hardware-efficient variational quantum eigensolver for small molecules and quantum magnets." Nature 549.7671 (2017): 242.


%\cite{Smith:2019mek}
\bibitem{Smith:2019mek}
A.~Smith, M.~S.~Kim, F.~Pollmann and J.~Knolle,
``Simulating quantum many-body dynamics on a current digital quantum computer,''
npj Quantum Inf. \textbf{5}, 106 (2019)
doi:10.1038/s41534-019-0217-0
[arXiv:1906.06343 [quant-ph]].

\bibitem{Miceli} R. Miceli, M. McGuigan "Quantum computation and visualization of Hamiltonians using discrete quantum mechanics and IBM QISKit", Proc. New York Scientific Data Summit, arXiv:1812.01044 [quant-ph] (2018). 

\bibitem{Okock} P. Okock, "A Matrix Method for Solving the Schrodinger Equation" Masters Thesis (2015).

\bibitem{Korsch} H. Korsch, M. Gluck, "Computing Eigenvalues Made Easy", Eur. J. Phys. 23 (2002).

\bibitem{Motycka} J. Motycka, J. Urban, P. Babinec, "Eigenvalues of supersymmetric matrix models", Act. Phys.Pol. B 45, 1743 (2014). 
%\cite{NASA/WMAP:2013} 
%\cite{Wiltshire:2003} 
%\cite{Valentino} 
%\cite{Brizuela:2016gnz}
%\cite{Garay:1990re}
%\cite{Halliwell:1986ja}
%\cite{Ha}
%\cite{Atkatz}
%\cite{deAlwis:2018sec}
%\cite{Domingos}

\bibitem{NASA/WMAP}
    NASA/WMAP Science Team “WMAP 9 Year Mission Results.” NASA, NASA, 2013, map.gsfc.nasa.gov/news/index.html.

%\cite{Wiltshire:2003} 
\bibitem{Wiltshire:2003}
D.~L.~Wiltshire,
``An Introduction to quantum cosmology,''
[arXiv:gr-qc/0101003 [gr-qc]].
  

%\cite{Valentino} 
\bibitem{Valentino}
E.~Di Valentino and L.~Mersini-Houghton,
``Testing Predictions of the Quantum Landscape Multiverse 1: The Starobinsky Inflationary Potential,''
JCAP \textbf{03}, 002 (2017)
doi:10.1088/1475-7516/2017/03/002
[arXiv:1612.09588 [astro-ph.CO]].
    
    
    
%\cite{Brizuela:2016gnz}
\bibitem{Brizuela:2016gnz}
D.~Brizuela, C.~Kiefer and M.~Kr\"amer,
``Quantum-gravitational effects on gauge-invariant scalar and tensor perturbations during inflation: The slow-roll approximation,''
Phys. Rev. D \textbf{94}, no.12, 123527 (2016)
doi:10.1103/PhysRevD.94.123527
[arXiv:1611.02932 [gr-qc]].


%\cite{Garay:1990re}
\bibitem{Garay:1990re} 
  L.~J.~Garay, J.~J.~Halliwell and G.~A.~Mena Marugan,
  ``Path integral quantum cosmology: A Class of exactly soluble scalar field minisuperspace models with exponential potentials,''
  Phys.\ Rev.\ D {\bf 43}, 2572 (1991).
  doi:10.1103/PhysRevD.43.2572
  %%CITATION = doi:10.1103/PhysRevD.43.2572;%%
  %56 citations counted in INSPIRE as of 10 Mar 2020

%\cite{Halliwell:1986ja}
\bibitem{Halliwell:1986ja} 
  J.~J.~Halliwell,
  ``Scalar Fields in Cosmology with an Exponential Potential,''
  Phys.\ Lett.\ B {\bf 185}, 341 (1987).
  doi:10.1016/0370-2693(87)91011-2
  %%CITATION = doi:10.1016/0370-2693(87)91011-2;%%
  %414 citations counted in INSPIRE as of 10 Mar 2020

%\cite{Ha}
\bibitem{Ha}
%\cite{Halliwell:1990uy}
J.~J.~Halliwell,
``INTRODUCTORY LECTURES ON QUANTUM COSMOLOGY,''
[arXiv:0909.2566 [gr-qc]].

%\cite{Atkatz}
\bibitem{Atkatz}
Atkatz, David. (1994). Quantum Cosmology for Pedestrians. American Journal of Physics - AMER J PHYS. 62. 619-627. 10.1119/1.17479. 

%\cite{deAlwis:2018sec}
\bibitem{deAlwis:2018sec}
S.~P.~de Alwis,
``Wave function of the Universe and CMB fluctuations,''
Phys. Rev. D \textbf{100}, no.4, 043544 (2019)
doi:10.1103/PhysRevD.100.043544
[arXiv:1811.12892 [hep-th]].

%\cite{Domingos}
%\bibitem{Domingos}
%Soares, Domingos. “Einstein's Static Universe.” ArXiv.org, 26 Mar. %2012,
%[arXiv:1203.4513 [physics.gen-ph]].

%\cite{Peebles}
%\cite{Peebles}
\bibitem{Peebles}
P.~J.~E.~Peebles and B.~Ratra,
``The Cosmological Constant and Dark Energy,''
Rev. Mod. Phys. \textbf{75}, 559-606 (2003)
doi:10.1103/RevModPhys.75.559
[arXiv:astro-ph/0207347 [astro-ph]].



%\cite{Maloney}
%\cite{Font:2002pq}
%\cite{Brown:2014sba}
%\cite{Brown:2013mwa}
%\cite{Brown:2013fba}
\bibitem{Maloney}
A.~Maloney, E.~Silverstein and A.~Strominger,
``De Sitter space in noncritical string theory,''
[arXiv:hep-th/0205316 [hep-th]].

%\cite{Font:2002pq}
\bibitem{Font:2002pq}
A.~Font and A.~Hernandez,
``Nonsupersymmetric orbifolds,''
Nucl. Phys. B \textbf{634}, 51-70 (2002)
doi:10.1016/S0550-3213(02)00336-X
[arXiv:hep-th/0202057 [hep-th]].

\bibitem{Brown:2014sba} 
  A.~R.~Brown, A.~Dahlen and A.~Masoumi,
  ``Flux compactifications on $(S_2)^N$,''
  Phys.\ Rev.\ D {\bf 90}, no. 4, 045016 (2014)
  doi:10.1103/PhysRevD.90.045016
  [arXiv:1401.7321 [hep-th]].
  %%CITATION = doi:10.1103/PhysRevD.90.045016;%%
  %6 citations counted in INSPIRE as of 26 Dec 2018

%\cite{Brown:2013mwa}
\bibitem{Brown:2013mwa} 
   A.~R.~Brown and A.~Dahlen,
  ``Spectrum and stability of compactifications on product manifolds,''
  Phys.\ Rev.\ D {\bf 90}, no. 4, 044047 (2014)
  doi:10.1103/PhysRevD.90.044047
  [arXiv:1310.6360 [hep-th]].
  %%CITATION = doi:10.1103/PhysRevD.90.044047;%%
  %11 citations counted in INSPIRE as of 26 Dec 2018

%\cite{Brown:2013fba}
\bibitem{Brown:2013fba} 
   A.~R.~Brown, A.~Dahlen and A.~Masoumi,
  ``Compactifying de Sitter space naturally selects a small cosmological constant,''
  Phys.\ Rev.\ D {\bf 90}, no. 12, 124048 (2014)
  doi:10.1103/PhysRevD.90.124048
  [arXiv:1311.2586 [hep-th]].
  %%CITATION = doi:10.1103/PhysRevD.90.124048;%%
  %6 citations counted in INSPIRE as of 26 Dec 2018




%\cite{Linde:1981zj}
%\cite{Frieman:1985xs}
%\cite{Kolb:1986nj}
%\cite{Kolb:1987dd}

\bibitem{Linde:1981zj}
A.~D.~Linde,
``Decay of the False Vacuum at Finite Temperature,''
Nucl. Phys. B \textbf{216}, 421 (1983)
[erratum: Nucl. Phys. B \textbf{223}, 544 (1983)]

%\cite{Frieman:1985xs}
\bibitem{Frieman:1985xs}
J.~A.~Frieman and E.~W.~Kolb,
``Semiclassical Instability of Compactification,''
Phys. Rev. Lett. \textbf{55}, 1435 (1985)
doi:10.1103/PhysRevLett.55.1435

%\cite{Kolb:1986nj}
\bibitem{Kolb:1986nj}
E.~W.~Kolb,
``Cosmology and Extra Dimensions,''
FERMILAB-PUB-86-138-A.

%\cite{Kolb:1987dd}
\bibitem{Kolb:1987dd}
E.~W.~Kolb,
``Vacuum leaks in extra dimensions,''
FERMILAB-PUB-87-104-A.


%\cite{Spergel:1999mh}
%\cite{Forestell:2016qhc}
%\cite{Faraggi:2000pv}
%\cite{McGuigan:2008pz}
\bibitem{Spergel:1999mh}
D.~N.~Spergel and P.~J.~Steinhardt,
``Observational evidence for selfinteracting cold dark matter,''
Phys. Rev. Lett. \textbf{84}, 3760-3763 (2000)
doi:10.1103/PhysRevLett.84.3760
[arXiv:astro-ph/9909386 [astro-ph]].

%\cite{Forestell:2016qhc}
\bibitem{Forestell:2016qhc}
L.~Forestell, D.~E.~Morrissey and K.~Sigurdson,
``Non-Abelian Dark Forces and the Relic Densities of Dark Glueballs,''
Phys. Rev. D \textbf{95}, no.1, 015032 (2017)
doi:10.1103/PhysRevD.95.015032
[arXiv:1605.08048 [hep-ph]].

%\cite{Faraggi:2000pv}
\bibitem{Faraggi:2000pv}
A.~E.~Faraggi and M.~Pospelov,
``Selfinteracting dark matter from the hidden heterotic string sector,''
Astropart. Phys. \textbf{16}, 451-461 (2002)
doi:10.1016/S0927-6505(01)00121-9
[arXiv:hep-ph/0008223 [hep-ph]].





%\cite{Obied}
%\bibitem{Obied}
%Obied, Georges, et al. “De Sitter Space and the Swampland.” %ArXiv.org, 17 July 2018, arxiv.org/abs/1806.08362. 



%\cite{Kapetanakis}
%\cite{Cavaglia:1996ek}

\bibitem{Kapetanakis}
D.~Kapetanakis, G.~Koutsoumbas, A.~Lukas and P.~Mayr,
``Quantum cosmology with Yang-Mills fields,''
Nucl. Phys. B \textbf{433}, 435-466 (1995)
doi:10.1016/0550-3213(94)00441-G
[arXiv:hep-th/9403131 [hep-th]].

%\cite{Cavaglia:1996ek}
\bibitem{Cavaglia:1996ek}
M.~Cavaglia,
``Quantization of gauge systems: application to minisuperspace models in canonical quantum gravity,''

%\cite{Maleknejad:2011jw}
\bibitem{Maleknejad:2011jw}
A.~Maleknejad and M.~M.~Sheikh-Jabbari,
``Gauge-flation: Inflation From Non-Abelian Gauge Fields,''
Phys. Lett. B \textbf{723}, 224-228 (2013)
doi:10.1016/j.physletb.2013.05.001
[arXiv:1102.1513 [hep-ph]].

%\cite{Wiltshire2}
%\bibitem{Wiltshire2}
%Wiltshire, D. L. “Wave Functions for Arbitrary Operator Ordering in the De Sitter Minisuperspace Approximation.” ArXiv.org, 24 May 1999, arxiv.org/abs/gr-qc/9905090. 

%\cite{Viglioni}
%\bibitem{Viglioni}
%Viglioni, Arthur, and Domingos Soares. “Note on the Classical Solutions of Friedmann's Equation.” ArXiv.org, 2 Dec. 2011, arxiv.org/abs/1007.0598. 

%\cite{Birrell:1982ix}
%\cite{Chitre:1977ip}
\bibitem{Birrell:1982ix}
N.~D.~Birrell and P.~C.~W.~Davies,
``Quantum Fields in Curved Space,''

%\cite{Chitre:1977ip}
\bibitem{Chitre:1977ip}
D.~M.~Chitre and J.~B.~Hartle,
``Path Integral Quantization and Cosmological Particle Production: An Example,''
Phys. Rev. D \textbf{16}, 251-260 (1977)
doi:10.1103/PhysRevD.16.251

%\cite{Zhang:2008jy}
\bibitem{Zhang:2008jy}
H.~H.~Zhang, K.~X.~Feng, S.~W.~Qiu, A.~Zhao and X.~S.~Li,
``On analytic formulas of Feynman propagators in position space,''
Chin. Phys. C \textbf{34}, 1576-1582 (2010)
doi:10.1088/1674-1137/34/10/005
[arXiv:0811.1261 [math-ph]].

%\cite{Brown:1990iv}
%\cite{Teitelboim:1980my}
%\cite{Henneaux:1981su}
%\cite{Ambjorn:1990jh}
%\cite{Greensite:1991qt}
%\cite{Garay:1990re}
%\cite{Halliwell:1986ja}
\bibitem{Brown:1990iv}
J.~D.~Brown and E.~A.~Martinez,
``Lorentzian Path Integral for Minisuperspace Cosmology,''
Phys. Rev. D \textbf{42}, 1931-1943 (1990)
doi:10.1103/PhysRevD.42.1931

%\cite{Teitelboim:1980my}
\bibitem{Teitelboim:1980my}
C.~Teitelboim,
``Proper Time Approach to the Quantization of the Gravitational Field,''
Phys. Lett. B \textbf{96}, 77-82 (1980)

%\cite{Henneaux:1981su}
\bibitem{Henneaux:1981su}
M.~Henneaux, M.~Pilati and C.~Teitelboim,
``Explicit Solution for the Zero Signature (Strong Coupling) Limit of the Propagation Amplitude in Quantum Gravity,''
Phys. Lett. B \textbf{110}, 123-128 (1982)

%\cite{Ambjorn:1990jh}
\bibitem{Ambjorn:1990jh}
J.~Ambjorn, J.~Greensite and S.~Varsted,
``A Nonperturbative definition of 2-D quantum gravity by the fifth time action,''
Phys. Lett. B \textbf{249}, 411-416 (1990)
doi:10.1016/0370-2693(90)91008-Y

%\cite{Greensite:1991qt}
\bibitem{Greensite:1991qt}
J.~Greensite,
``A Fifth time action approach to the conformal instability problem in Euclidean quantum gravity,''
Nucl. Phys. B \textbf{361}, 729-746 (1991)
doi:10.1016/0550-3213(91)90602-T

%\cite{Garay:1990re}
\bibitem{Garay:1990re}
L.~J.~Garay, J.~J.~Halliwell and G.~A.~Mena Marugan,
``Path integral quantum cosmology: A Class of exactly soluble scalar field minisuperspace models with exponential potentials,''
Phys. Rev. D \textbf{43}, 2572-2589 (1991)

%\cite{Halliwell:1986ja}
\bibitem{Halliwell:1986ja}
J.~J.~Halliwell,
``Scalar Fields in Cosmology with an Exponential Potential,''
Phys. Lett. B \textbf{185}, 341 (1987)





%\cite{Waldron:2004gg}
\bibitem{Waldron:2004gg}
A.~Waldron,
``Milne and torus universes meet,''
[arXiv:hep-th/0408088 [hep-th]].

%\cite{Halliwell:1989pu}
%\cite{Duru:1984dx}
\bibitem{Halliwell:1989pu}
J.~J.~Halliwell and R.~C.~Myers,
``Multiple Sphere Configurations in the Path Integral Representation of the Wave Function of the Universe,''
Phys. Rev. D \textbf{40}, 4011 (1989)
doi:10.1103/PhysRevD.40.4011


%\cite{Duru:1984dx}
\bibitem{Duru:1984dx}
I.~H.~Duru,
``MORSE POTENTIAL GREEN'S FUNCTION WITH PATH INTEGRALS,''
Phys. Rev. D \textbf{28}, 2689-2692 (1983)
doi:10.1103/PhysRevD.28.2689

\bibitem{Kramer} T. Kramer, M. Moshinsky, "Tunneling out of a time dependent well", arXiv:quant-ph/0505099 (2005).



%\cite{Vilenkin:1987kf}%\cite{Gasperini:2021eri}
\bibitem{Vilenkin:1987kf}
A.~Vilenkin,
``Quantum Cosmology and the Initial State of the Universe,''
Phys. Rev. D \textbf{37}, 888 (1988)
doi:10.1103/PhysRevD.37.888


%\cite{Gasperini:2021eri}
\bibitem{Gasperini:2021eri}
M.~Gasperini,
``Quantum string cosmology,''
Universe \textbf{7}, 14 (2021)
[arXiv:2101.01070 [gr-qc]].


%\cite{Casali:2021ewu}
%\cite{Betzios:2020nry}
%\cite{Iliesiu:2020zld}
%\cite{Martinec:1984fs}
%\cite{Witten:2020wvy}
%\cite{Maldacena:2005he}
%\cite{Douglas:2003up}
\bibitem{Casali:2021ewu}
E.~Casali, D.~Marolf, H.~Maxfield and M.~Rangamani,
``Baby Universes and Worldline Field Theories,''
[arXiv:2101.12221 [hep-th]].

%\cite{Betzios:2020nry}
\bibitem{Betzios:2020nry}
P.~Betzios and O.~Papadoulaki,
``Liouville theory and Matrix models: A Wheeler DeWitt perspective,''
JHEP \textbf{09}, 125 (2020)
%doi:10.1007/JHEP09(2020)125
[arXiv:2004.00002 [hep-th]].

%\cite{Iliesiu:2020zld}
\bibitem{Iliesiu:2020zld}
L.~V.~Iliesiu, J.~Kruthoff, G.~J.~Turiaci and H.~Verlinde,
``JT gravity at finite cutoff,''
SciPost Phys. \textbf{9}, 023 (2020)
%doi:10.21468/SciPostPhys.9.2.023
[arXiv:2004.07242 [hep-th]].





 
 %\cite{Martinec:1984fs}
\bibitem{Martinec:1984fs}
E.~J.~Martinec,
``Soluble Systems in Quantum Gravity,''
Phys. Rev. D \textbf{30}, 1198 (1984)
doi:10.1103/PhysRevD.30.1198

%\cite{Witten:2020wvy}
\bibitem{Witten:2020wvy}
E.~Witten,
``Matrix Models and Deformations of JT Gravity,''
Proc. Roy. Soc. Lond. A \textbf{476}, no.2244, 20200582 (2020)
doi:10.1098/rspa.2020.0582
[arXiv:2006.13414 [hep-th]].



%\cite{Maldacena:2005he}
\bibitem{Maldacena:2005he}
J.~M.~Maldacena and N.~Seiberg,
``Flux-vacua in two dimensional string theory,''
JHEP \textbf{09}, 077 (2005)
doi:10.1088/1126-6708/2005/09/077
[arXiv:hep-th/0506141 [hep-th]].


%\cite{Douglas:2003up}
\bibitem{Douglas:2003up}
M.~R.~Douglas, I.~R.~Klebanov, D.~Kutasov, J.~M.~Maldacena, E.~J.~Martinec and N.~Seiberg,
``A New hat for the c=1 matrix model,''
[arXiv:hep-th/0307195 [hep-th]].







%\cite{Berger:1993fm}
%\cite{Misner:1973zz}
%\cite{FernandoBarbero:2010qy}
%\cite{McGuigan:1990nd}
%\cite{Fischler:1989se}
%\cite{Berger:1972pg}
%\cite{Berger:1975kn}




%\cite{Misner:1973zz}
\bibitem{Misner:1973zz}
C.~W.~Misner,
``A Minisuperspace Example: The Gowdy T3 Cosmology,''
Phys. Rev. D \textbf{8}, 3271-3285 (1973)
doi:10.1103/PhysRevD.8.3271

%\cite{FernandoBarbero:2010qy}
\bibitem{FernandoBarbero:2010qy}
J.~F.~Barbero G. and E.~J.~S.~Villasenor,
``Quantization of Midisuperspace Models,''
Living Rev. Rel. \textbf{13}, 6 (2010)
doi:10.12942/lrr-2010-6
[arXiv:1010.1637 [gr-qc]].

%\cite{McGuigan:1990nd}
\bibitem{McGuigan:1990nd}
M.~McGuigan,
``The Gowdy cosmology and two-dimensional gravity,''
Phys. Rev. D \textbf{43}, 1199-1211 (1991)

%\cite{Fischler:1989se}
\bibitem{Fischler:1989se}
W.~Fischler, D.~Morgan and J.~Polchinski,
``Quantum Nucleation of False Vacuum Bubbles,''
Phys. Rev. D \textbf{41}, 2638 (1990)
doi:10.1103/PhysRevD.41.2638

%\cite{Berger:1972pg}
\bibitem{Berger:1972pg}
B.~K.~Berger, D.~M.~Chitre, V.~E.~Moncrief and Y.~Nutku,
``Hamiltonian formulation of spherically symmetric gravitational fields,''
Phys. Rev. D \textbf{5}, 2467-2470 (1972)
doi:10.1103/PhysRevD.5.2467

%\cite{Berger:1975kn}
\bibitem{Berger:1975kn}
B.~K.~Berger,
``Quantum Cosmology: Exact Solution for the Gowdy T**3 Model,''
Phys. Rev. D \textbf{11}, 2770-2780 (1975)
doi:10.1103/PhysRevD.11.2770

%\cite{Feldbrugge:2017kzv}
%\cite{Caputa:2018asc}
\bibitem{Feldbrugge:2017kzv}
J.~Feldbrugge, J.~L.~Lehners and N.~Turok,
``Lorentzian Quantum Cosmology,''
Phys. Rev. D \textbf{95}, no.10, 103508 (2017)
doi:10.1103/PhysRevD.95.103508
[arXiv:1703.02076 [hep-th]].

%\cite{Caputa:2018asc}
\bibitem{Caputa:2018asc}
P.~Caputa and S.~Hirano,
``Airy Function and 4d Quantum Gravity,''
JHEP \textbf{06}, 106 (2018)
doi:10.1007/JHEP06(2018)106
[arXiv:1804.00942 [hep-th]].

%\cite{McGuigan:1993ma}
\bibitem{McGuigan:1993ma}
M.~McGuigan,
``Third quantization and black holes,''
[arXiv:hep-th/9212044 [hep-th]].


  
   \end{thebibliography}
\end{document}